\documentclass[manuscript,screen,final]{acmart}

\renewcommand\footnotetextcopyrightpermission[1]{} 
\AtBeginDocument{%
  }

\setcopyright{acmlicensed}
\copyrightyear{2024}
\acmYear{2024}
\acmDOI{XXXXXXX.XXXXXXX}

\acmISBN{978-1-4503-XXXX-X/18/06}

\usepackage{amsmath,amssymb,amsfonts}
\usepackage{algorithmic}
\usepackage{graphicx}
\usepackage{textcomp}
\usepackage{xcolor}

\usepackage{mathtools}
\usepackage{amsthm}
\usepackage{threeparttable}
\usepackage{natbib}
\setcitestyle{numbers,square}
\usepackage[capitalize,noabbrev]{cleveref}
\theoremstyle{plain}

\theoremstyle{definition}

\theoremstyle{remark}

\usepackage[algo2e,ruled,vlined]{algorithm2e}
\usepackage{algorithmic}
\usepackage{booktabs}
\usepackage{threeparttable}
\usepackage{multirow}
\usepackage{booktabs}
\usepackage{algorithm}
\usepackage{algorithmic}
\usepackage{bm}
\usepackage{colortbl}
\usepackage{url}
\usepackage{fancyhdr}
\newcommand{\wbnote}[1]{\textbf{\color{red}#1}}

\begin{document}

\title{One Backpropagation in Two Tower Recommendation Models}

\renewcommand\footnotetextcopyrightpermission[1]{}

\author{Erjia Chen}
\affiliation{%
  \institution{Huazhong University of Science and Technology}
  \city{Wuhan}
  \country{China}}
\email{larst@affiliation.org}

\author{Bang Wang}
\affiliation{%
	\institution{Huazhong University of Science and Technology}
	\city{Wuhan}
	\country{China}}
\email{larst@affiliation.org}

\renewcommand{\shortauthors}{Erjia Chen, Bang Wang}

\begin{abstract}
	Recent years have witnessed extensive researches on developing two tower recommendation models for relieving information overload. Four building modules can be identified in such models, namely, user-item encoding, negative sampling, loss computing and backpropagation updating. To the best of our knowledge, existing algorithms have researched only on the first three modules, yet neglecting the backpropagation module. They all adopt a kind of two backpropagation strategy, which are based on an implicit assumption of equally treating users and items in the training phase.
	
	\par
	In this paper, we challenge such an equal training assumption and propose a novel one backpropagation updating strategy, which keeps the normal gradient backpropagation for the item encoding tower, but cuts off the backpropagation for the user encoding tower. Instead, we propose a moving-aggregation updating strategy to update a user encoding in each training epoch. Except the proposed backpropagation updating module, we implement the other three modules with the most straightforward choices. Experiments on four public datasets validate the effectiveness and efficiency of our model in terms of improved recommendation performance and reduced computation overload over the state-of-the-art competitors.
\end{abstract}



\maketitle

\section{Introduction}
Recommender systems have been developed for decades to help people dealing with the issue of information overload by recommending a user only a few of his/her mostly interested items~\cite{Aljukhadar:2012:IJEC,Aljukhadar:2010:RecSys}. With the increasing concern on privacy, users and items are only equipped with unique identifiers without more descriptions, and various users behaviors, such as view, click, purchase and etc., are simplified as binary interactions between users and items, i.e., interacted or uninteracted. Recommendation only based on such binary interactions is also well known as the \textit{One Class Collaborative Filtering} (OCCF)~\cite{Pan:2008:IEEE,Hu:2008:IEEE}: How to recommend a sorted list consisting of uninteracted items for each user based only on a user-item interaction matrix?

\par
Extensive research efforts have been devoted to dealing with the OCCF problem~\cite{Hsieh:2017:WWW,Tran:2019:SIGIR,Liu:2023:ICDE,Xiangnan:2020:SIGIR,Liang:2018:WWW,Chae:2018:CIKM,Li:2022:WWW,Bao:2022:NIPS,Bao:2022:IEEE,Lee:2021:SIGIR,Li:2020:AAAI,Wei:2023:SIGIR}. The key to success lies in how to learn high quality users' and items' representations from their unique identifiers and binary interactions. A user-item predicted rating matrix can then be computed based on users' and items' representations to produce recommendation lists. Most proposed algorithms for the OCCF problem have adopted a kind of \textit{two tower recommendation model}\wbnote~\cite{Mao:2021:CIKM,Huang:2013:CIKM,Yang:2020:WWW,Su:2023:SIGIR}, in which one tower for encoding users' representations and another for encoding items' representations. Furthermore, these algorithms have employed the \textit{contrastive learning} technique~\cite{Chen:2020:ICML,Chuang:2020:NIPS,Khosla:2020:NIPS,Song:2020:NIPS} to differentiate interacted and uninteracted items in encoding representations.

\par
We can identify four building modules in most two tower recommendation models, namely, \textit{user-item encoding}, \textit{negative sampling}, \textit{loss computing} and \textit{backpropagation updating}. Many recommendation algorithms have been proposed with the focus of improving the first three modules. For user-item encoding, many advanced yet sophisticated neural networks have been designed with the aim of better encoding collaborative information from the interaction matrix~\cite{Xiangnan:2017:WWW, Wang:2019:SIGIR,Xiangnan:2020:SIGIR,Liang:2018:WWW,Chae:2018:CIKM,Li:2022:WWW}. For negative sampling~\cite{Chen:2017:KDD,Steffen:2009:UAI,Zhang:2013:SIGIR,Hsieh:2017:WWW,Tran:2019:SIGIR,Liu:2023:ICDE,Chen:2023:WWW}, the main objective is to select those true negatives from uninteracted, i.e., an uninteracted item truly disliked by a user. For loss computing, the InfoNCE~\cite{Oord:2018:arxiv} has become a rather popular contrastive learning loss function, and some algorithms have proposed to further improve it, say for example, by using different similarity computation functions~\cite{Hsieh:2017:WWW,Koren:2009:Computer}.

\par
To the best of our knowledge, all the two tower recommendation models have adopted the same model training strategy: The loss gradient is backpropagated into both the user tower and item tower for each to update respective encodings. We call this training strategy as \textit{two backpropagation} in this paper, which is a kind of symmetric backpropagation strategy based on \textit{an implicit assumption of treating users and items equally in model training}. We agree that the two backpropagation strategy is a straightforward choice, as the loss is computed over training users' and items' interactions. However, we argue that it may not be a wise choice in a two tower model for solving the OCCF problem.

\par
In this paper, we challenge such an equal treating assumption in two backpropagation for model training. We first argue that even though items are equipped with only identifiers, they are indeed also with some intrinsic attributes (like a movie genre and director), even not exposed in the OCCF scenarios. We next argue that a collaborative filtering model could be capable of learning such unexposed attributes as encoded in the learned items' representations, and as such, items can be clustered into \textit{latent types}. We also argue that a user's interests could be diverse, embodied in that his/her interacted items normally are distributed in different latent types. We note that in the batch-based training phase, the gradient direction would be dominated by those items with a same type. As such, using the two backpropagation may lead to the learned representation of a user too much favoring those of the same type items, while neglecting items of other types.

\begin{figure*}[t]
	\includegraphics[width=\textwidth]{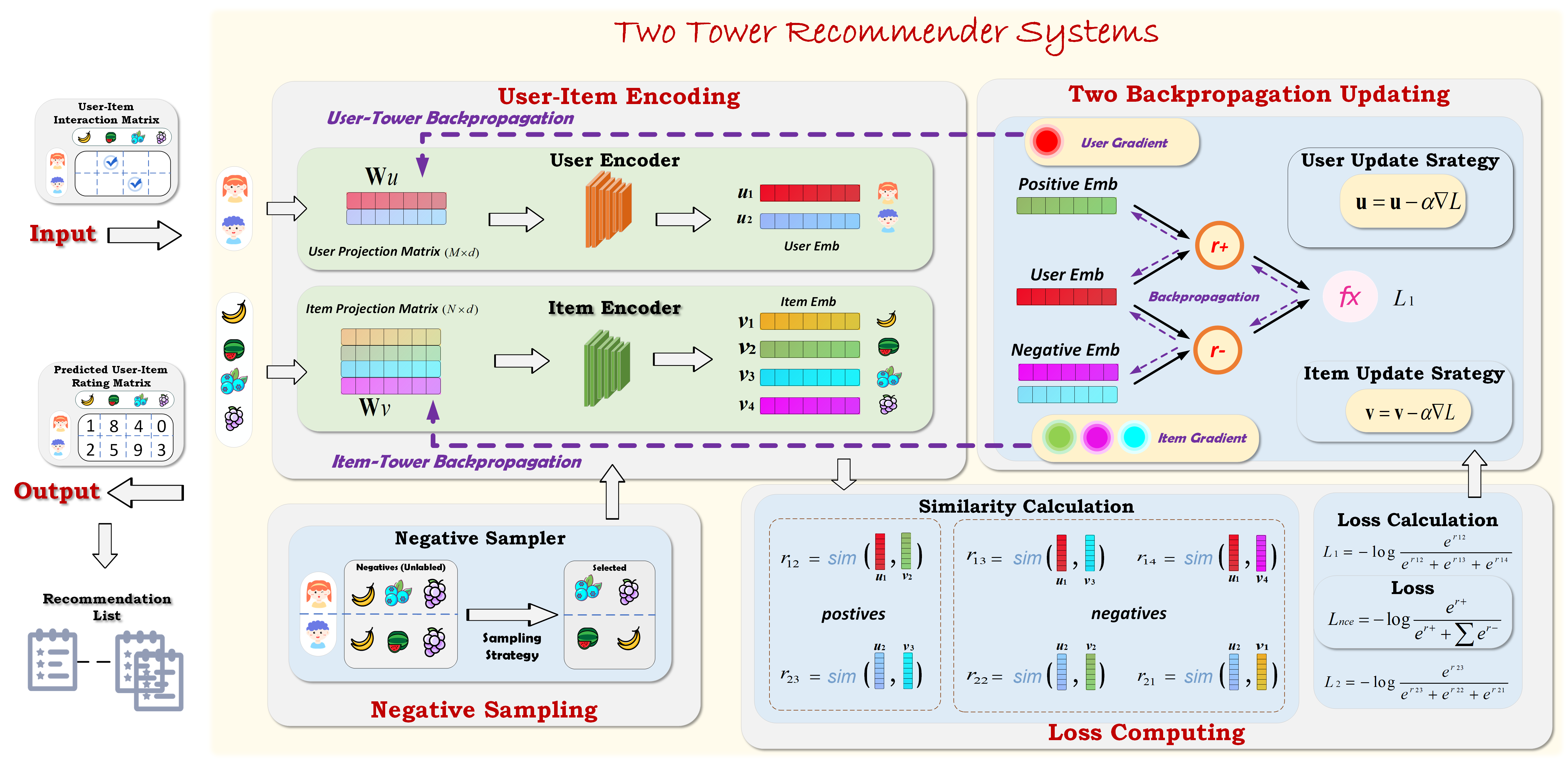}
	\caption{Illustration of a typical two tower recommendation model with four building modules, namely, user-item encoding, negative sampling, loss computing, and backpropagation updating.}
	\label{Fig:TwoTowerModel}
\end{figure*}

\par
Motivated from such considerations, we propose \textit{one backpropagation} (OneBP) in two tower recommendation models. The core idea is rather simple: We keep the normal gradient backpropagation for the item encoding tower, but cut off the backpropagation for the user encoding tower. Instead, we propose a \textit{moving-aggregation} strategy to update a user encoding in each training epoch. In particular, the user encoding is updated by the aggregation over his/her previous encoding and his/her just-interacted item's encoding. Except the new backpropagation strategy, we implement other three modules with the most straightforward choices, namely, the projection matrix-based user-item encoding, random negative sampling and plain InfoNCE loss. We experiment our OneBP model on four widely used public datasets. Results validate its superiority over representative state-of-the-art models in terms of large improvement margins on recommendation performance, as well as its computation efficiency.

\par
The rest of the paper is organized as follows: Section~\ref{Sec:TwoTowerModel} introduces the two tower model, and the related works are provided in Section~\ref{Sec:RelatedWork}. Our OneBP model is presented in Section~\ref{Sec:OneBPModel} and experimented in Section~\ref{Sec:Experiments}. The paper is concluded in Section~\ref{Sec:Conclusion}.

\section{Two Tower Recommendation Models}\label{Sec:TwoTowerModel}
\textbf{Preliminary: }
The \textit{one-class collaborative filtering} (OCCF) for recommendation can be defined as the following problem. Let $\mathcal{U}=\{u_i\}_1^M$ denote the user set, and $\mathcal{V}=\{v_j\}_1^N$ the item set. We note that in the OCCF problem setting, each user (each item) is only equipped with a unique identifier, yet without any other side information. Let $a_{ij}=1$ denote an interaction (such as click, purchase) that had happened between $u_i$ and $v_j$; Otherwise, $a_{ij}$ is not defined. The input of an OCCF recommender is the \textit{user-item interaction matrix} $\mathbf{I}=[a_{ij}]_{M\times N}$; The output is a predicted \textit{user-item rating matrix} $\mathbf{R}=[r_{ij}]_{M \times N}$ with each $r_{ij}$ being a real-valued prediction. Based on $\mathbf{R}$, a ranking list for recommending top-$K$ un-interacted items can be generated for each user.

\par
\textbf{Two Tower Recommendation Model: }
The two tower model has been widely adopted to solve the OCCF problem for recommendation, which, based on the interaction matrix, is to encode users and items from identifiers to representations for their similarity computation and ranking list generation. It is called "two tower" in that it consists of two parallel and independent encoders: One for encoding users, and another for encoding items. For the success of contrastive learning, recent two tower models often employ \textit{negative sampling} and \textit{contrastive loss computing} for training the two encoders via \textit{error backpropagation}. We can identify four building modules of a two tower recommendation model, namely, \textit{user-item encoding}, \textit{negative sampling}, \textit{loss computing}, and \textit{backpropagation updating}, as illustrated in Fig.~\ref{Fig:TwoTowerModel}.

\textbf{User-Item Encoding: }
This module consists of a user encoder to encode a user from its one-hot $M$-dimensional identifier $u_{i}$ to a real-valued $d$-dimensional ($d\ll M$) representation $\mathbf{u}_i$, and an item encoder to encode an item $v_j$ to a $d$-dimensional representation $\mathbf{v}_j$:
\begin{align}
	\mathbf{u}_i = UserEncoder(u_{i}) , \nonumber
	\mathbf{v}_j = ItemEncoder(v_{j}). \nonumber
\end{align}
There are many choices for the encoders, such as the widely used \textit{matrix factorization} (MF)~\cite{Koren:2009:Computer} and recently proposed neural networks-based encoders like the NGCF~\cite{Wang:2019:SIGIR} and LightGCN~\cite{Xiangnan:2020:SIGIR}.

\textbf{Negative Sampling: }
Due to the nature of binary interactions, the \textit{contrastive learning} has been serving as the main solution approach for the OCCF problem. In the OCCF setting, a \textit{positive} sample is from interactions $(u_i, v_i^+)$ between a user $u_i$ and one of his/her interacted items; While a \textit{negative} sample $(u_i, v_i^-)$ is from one of his/her uninteracted items. In recommendation systems, uninteracted items are much more than interacted ones, and \textit{negative sampling} is needed to select a number of uninteracted items (say $N_s \ll N$):
\begin{equation}
	\mathcal{V}_{u_i}^- = NegativeSampling(\mathcal{V}, N_s)
\end{equation}
Notice that an uninteracted item could be the one not seen before by a user but not an uninterested item to this user. As such, many negative sampling algorithms have been proposed to select high-quality uninteracted items as negatives, such as the PNS~\cite{Mikolov:2013:NIPS,Chen:2017:KDD,Tang:2015:WWW}, DNS~\cite{Zhang:2013:SIGIR} and etc.

\textbf{Loss Computing: }
The contrastive loss computation can be further divided into two steps: \textit{similarity calculation} and \textit{loss calculation}. The former is to compute the predicted \textit{rating score} $r_{ij}$ between a user and an item:
\[
r_{ij} = Similarity\left(\bm{u_i},\bm{v_j}\right).
\]
The commonly used similarity function includes cosine, inner product function. Let $r_{ij}^+$ ($r_{ik}^-$) denote the rating between a user $u_i$ and his/her positive item $v_j$ (negative item $v_k$ sampled from his/her uninteracted item sets $\mathcal{V}_{u_i}^-$). In the OCCF problem, the popular contrastive loss function is the following InfoNCE loss~\cite{Oord:2018:arxiv}:
\begin{equation}\label{Eq:InfoNCE}	
	\mathcal{L} = - \sum_i \log \left( \frac{\exp(r_{ij}^+) }{\exp(r_{ij}^+) + \sum_{v_k \in \mathcal{V}_{u_i}^-} \exp({r_{ik}^-})} \right)
\end{equation}
We note that the pairwise loss function BPR~\cite{Steffen:2009:UAI} is a special case of InfoNCE using only one negative.

\par
\textbf{Backpropagation Updating:}
During the model training, the most used technique is the gradient backpropagation, which applies the gradients calculated from the loss to update the inputs of the encoders. To the best of our knowledge, all existing two tower models use the same updating strategy to update both the user encoder and item encoder. Although different encoders have different structures and parameters, the objective of gradient backpropagation is to update the user representation and item representation after encoders in each iteration. As such, we can summarize the backpropagation updating in one training epoch as follows:
\begin{equation}\label{Eq:TwoBPUser}
	\mathbf{u}_i \leftarrow \mathbf{u}_j - \alpha \nabla{\mathcal{L}}
\end{equation}
\begin{equation}\label{Eq:TwoBPItem}
	\mathbf{v}_j \leftarrow \mathbf{v}_j - \alpha \nabla{\mathcal{L}}
\end{equation}
where $\alpha$ is the learning rate and $\nabla{\mathcal{L}}$ is a partial derivative w.r.t. the parameters being updated.

\section{Related Work}\label{Sec:RelatedWork}
Extensive solutions have been proposed to deal with the OCCF problem. Yet to the best of our knowledge, all these solutions have focused only on the first three modules of the two tower model. Although some solutions involve with more than one module, we classify them into three categories according to their main focus.

\subsection{Encoder-based method}
For encoding users and items, many advanced encoders have been designed with the objective of learning more informative and accurate representations. The \textbf{MF} (Matrix Factorization) is the classic yet efficient encoder in recommendation systems~\cite{Koren:2009:Computer}, which iteratively updates two projection matrices $\mathbf{W}_u$ and $\mathbf{W}_v$ to encode users' and items' identifiers into representations.

\par
Recently, many advanced neural networks have been designed as encoders~\cite{Xiangnan:2017:WWW, Wang:2019:SIGIR,Xiangnan:2020:SIGIR,Liang:2018:WWW,Chae:2018:CIKM,Li:2022:WWW}, among which the NGCF~\cite{Wang:2019:SIGIR} and LightGCN~\cite{Xiangnan:2020:SIGIR} are two representative ones of discriminative model. The \textbf{NGCF}~\cite{Wang:2019:SIGIR} applies a graph neural network to recursively propagates and aggregates collaborative signals in the user-item interaction graph in order to capture high-order structural information. The \textbf{LightGCN}~\cite{Xiangnan:2020:SIGIR} argues that the feature transformation and non-linear activation in NGCF do not positively impact collaborative filtering. As such, the LightGCN significantly simplifies the graph neural network by only retaining the most crucial components, leading to improved performance.

\subsection{Negative Sampling method}
For negative sampling, the challenge is from the nature of unlabeled data in the OCCF problem. Many approaches have focused on selecting high-quality true negatives~\cite{Steffen:2009:UAI,Zhang:2013:SIGIR,Hsieh:2017:WWW,Tran:2019:SIGIR,Liu:2023:ICDE}, yet some focus on other criteria in negative sampling, such as mitigating popularity bias~\cite{Chen:2023:WWW,Mikolov:2013:NIPS,Liu:2023:arxiv}. In negative sampling, the probability of selecting an uninteracted item can be predefined (static) or adjusted in each iteration (dynamic).

\par
The \textbf{RNS} (random negative sampling)~\cite{Steffen:2009:UAI,Xiangnan:2020:SIGIR,Yu:2018:CIKM} is the widely used static negative sampling, which selects uninteracted items with equal probabilities. The \textbf{PNS} (popularity-based negative sampling)~\cite{Mikolov:2013:NIPS,Chen:2017:KDD,Tang:2015:WWW} computes the sampling probability for an uninteracted item to a user proportional to the item popularity that is defined as the interaction frequency of this item across all users in the training dataset.

\par
The \textbf{DNS} (Dynamic Negative Sampling)~\cite{Zhang:2013:SIGIR} is the widely used dynamic negative sampling algorithm, which assigns higher sampling probabilities to items with elevated encoding dot product scores obtained during each training iteration. The \textbf{HarS}~\cite{Hsieh:2017:WWW} is also a dynamic negative sampling strategy that leverages model information. It discerns items from the candidate set based on the Euclidean distance computation between users and items. Their selection process prioritizes items proximate to the anchor user for model training. The samples chosen through this method are characterized by a wealth of informative content.

\par
The \textbf{2stS}~\cite{Tran:2019:SIGIR} combines static and dynamic sampling strategy: In each iteration, it first utilizes the static popularity-based sampling probabilities to select a few negatives as a candidate set, and then applies encoding similarity information to select informative negatives from the candidate set.

\subsection{Loss Function}
Although the InfoNCE in Eq.~\eqref{Eq:InfoNCE} has been becoming a kind of standard loss function for contrastive learning, some have proposed to further improve it from two viewpoints: One is to modify the similarity calculation function~\cite{Hsieh:2017:WWW,Bao:2022:NIPS,Bao:2022:IEEE,Lee:2021:SIGIR,Li:2020:AAAI,Wei:2023:SIGIR}, and another is to alleviate the adverse impact of false negatives)~\cite{Chuang:2020:NIPS,Robinson:2020:ICLR}.

\par
The \textbf{CML} (Collaborative Metric Learning)~\cite{Hsieh:2017:WWW} addresses the limitation of the dot product in capturing relationships between similar entities by utilizing Euclidean distance instead of inner product. The \textbf{DPCML}~\cite{Bao:2022:NIPS} builds upon the CML by optimizing the representation of each user, allowing multiple representations for each user. During training process, DPCML selects samples with the shortest Euclidean distance to item encodings to simulate the diversity of user interests in reality. The \textbf{SFCML}~\cite{Bao:2022:IEEE} identifies bias introduced by the negative sampling module in the CML framework and constructs a CML architecture that eliminates the need for negative sampling while also reducing computational burdens.

\par
The \textbf{DCL} (Debiased Collaborative Filtering)~\cite{Chuang:2020:NIPS} addresses the issue of false negatives by correcting the scoring part in the loss function. The \textbf{HCL} (Hard Collaborative Learning)~\cite{Robinson:2020:ICLR} builds upon the DCL by further incorporating a hard negative sampling module.

\par

In addition to widely adopted methods such as InfoNCE and BPR, some models also design loss functions based on their own optimization objectives. For exmaple \textbf{BUIR}~\cite{Lee:2021:SIGIR} introduces a two-network architecture, where one network predicts the output of another. In the loss computation step, it uses cosine similarity to capture the relationships between users and items across the two networks and combining the negation of two similarity components and adding them together to form the loss function.

\section{The Proposed OneBP Method}\label{Sec:OneBPModel}
We adopt the two tower model for personalized recommendation yet with our new backpropagation strategy for updating users' and items' representations, that is, our focus is on the last module of backpropgation updating. For the other modules, we conservatively select the commonly used yet simple techniques without involving other complications.

\subsection{The common modules}
\textbf{Projection-based User-Item Encoding: } We use a very simple projection for encoding users and items. Let $\mathbf{W}_u \in \mathbb{R}^{M\times k}$ denote the user projection matrix which converts a user one-hot identifier vector into a $d$-dimensional real-valued representation. Simply put, the $i$-th row of $\mathbf{W}_u$ is the user $u_i$'s representation. Similarly, $\mathbf{W}_v \in \mathbb{R}^{N\times k}$ is the item projection matrix. The two projection matrices $\mathbf{W}_u$ and $\mathbf{W}_v$ are subject to be updated in the model training phase. Before model training, they are randomly initialized.

\textbf{Random Negative Sampling:} We adopt the most straightforward \textit{random negative sampling}: For each positive sample $(u_i, v_i^+)$, we randomly select his/her $N_s$ uninteracted items as negative samples $(u_i, v_i^-)$. We notice that although many advanced negative sampling techniques can be used to promote contrastive learning, the choice of the random one is with a significant merit of fast computation.

\textbf{InfoNCE Loss Computing: } We adopt the commonly used InfoNCE loss function Eq.~\eqref{Eq:InfoNCE} and use the dot product for similarity calculation.

\begin{figure}[t]
	\includegraphics[width=0.85\textwidth,height=0.4\textheight]{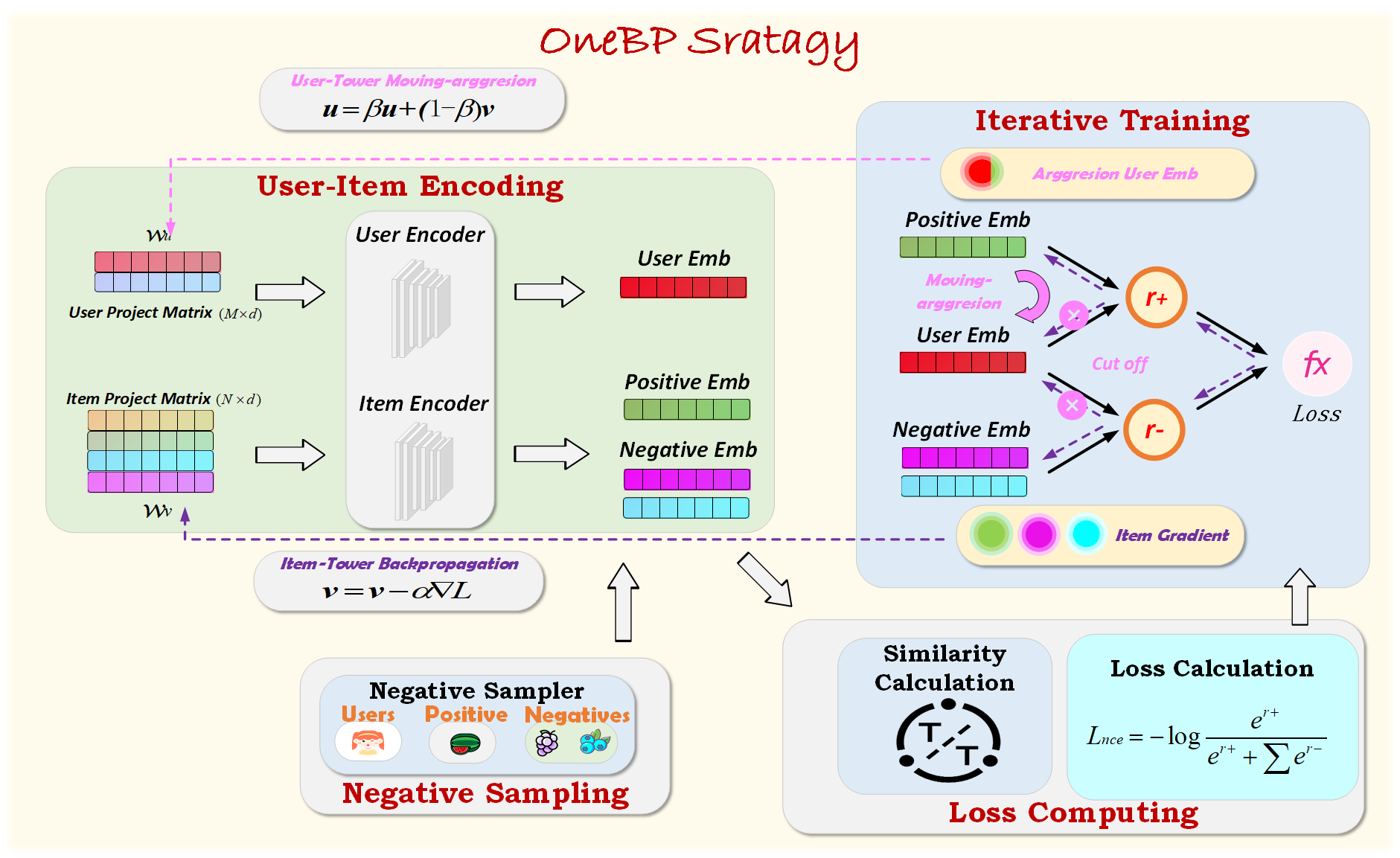}
	\caption{The proposed OneBP model: It
		cuts off the gradient backpropagation to the user encoder and uses moving-aggregation for user encoding update.}\label{Fig:OneBPModel}
\end{figure}

\subsection{One Backpropagation Updating}
As discussed in previous sections, existing two tower models employ the same loss gradient backpropagation strategy for updating both users' and items' representations, see Eq.~\eqref{Eq:TwoBPUser} and Eq.~\eqref{Eq:TwoBPItem}. We call this kind of updating strategy as \textit{two backpropagation}.

\par
In this paper, we propose to use loss gradient backpropagation for updating items' representation only, and propose to update users' representations by our moving-aggregation strategy. We call our updating strategy as "OneBP" for brevity. Fig.~\ref{Fig:OneBPModel} compares our OneBP with the two backpropagation. The core idea is to cut off the gradient flow towards users' encodings.

\par
We detail the execution of our OneBP strategy for one user in one training epoch as follows: Let $(u_i, v_j, \{v_k\})$ denote the selected user $u_i$ with his/her interacted item $v_j$ as positive and his/her uninteracted items $\{v_k\}$ as negatives. Our OneBP updates $\mathbf{u}_i$, $\mathbf{v}_j$, and $\mathbf{v}_k$ as follows:
\begin{equation}\label{Eq:OneBPItem}
	\mathbf{v}_j^{(t+1)} = \mathbf{v}_j^{(t)} - \alpha \nabla{\mathcal{L}} \;, \quad
	\mathbf{v}_k^{(t+1)} = \mathbf{v}_k^{(t)} - \alpha \nabla{\mathcal{L}}
\end{equation}
\begin{equation}\label{Eq:OneBPUser}
	\mathbf{u}_i^{(t+1)} = \beta \mathbf{u}_i^{(t)} + (1 - \beta) \mathbf{v}_j^{(t+1)}.
\end{equation}
Compared with the two backpropagation, our OneBP does not use gradient backpropagation for updating users' representations. Instead, Eq.~\eqref{Eq:OneBPUser} indicates that the user encoding is updated by the aggregation over his/her previous one and his/her just-updated interacted item's encoding in a training epoch.

\subsection{Discussions}\label{Sec:Discussion}
\subsubsection{Symmetric or Asymmetric} \label{Sec:Symmetric}
Obviously, existing two backpropagation performs a kind of symmetric encoding update for users and items. This implicitly indicates to treat users and items equally in the two tower model. On the contrary, our OneBP performs a kind of asymmetric update, which suggests to treat users and items differently. We argue our design philosophy as follows.

\par
We first argue that an item, whether or not it has been interacted by which users and how many users, possesses its intrinsic attributes (such as a banana is a kind of fruit and with yellow color). Although we only have items' identifers at first, their encodings after the training of a two tower recommendation model might be with a kind of \textit{latent types}. Fig.~\ref{Fig:EncodingClustering} illustrates the clustering phenomenon based on items' encodings. One cluster can be regarded as one latent type of items.

\par
We also argue that the interests of a user is often diverse, which can be observed from Fig.~\ref{Fig:EncodingClustering}, where the interacted items by a user in the training dataset spread in several clusters. However, the distribution of such interacted items may be unbalanced in different clusters, normally \textit{majority items} in one cluster and \textit{minority items} in other clusters. We note that in the batch-based training phase, the gradient direction would be dominated by those majority items. As such, using the two backpropagation may lead to the learned representation of a user too much favoring those of majority items, while neglecting minority items for learning a user's diverse interests. In our OneBP, we exploit each interacted item to independently update the user encoding by the moving-aggregation operation, so it could well learn attributes from both majority and minority items to enhance the capability of learning diverse interests for a user.

\subsubsection{Decoupling the Impact of False Negative Samples} We note that although contrastive learning is a powerful technique in many applications, it faces the challenge of \textit{false negative sample} in the OCCF problem. That is, an unlabeled item is a user interested one, but it is not interacted simply due to not yet exposed to this user. If such an unlabeled item is selected as a negative sample for contrastive loss computation, it can lead to update user encoding in an incorrect direction. This issue has been extensively studied in the literature~\cite{Rottmann:2020:EDA,Li:2023:ACM,Chen:2021:ICLR,Ding:2020:NIPS}. Although many algorithms have been proposed to alleviate the adverse impact of false negative samples~\cite{Chuang:2020:NIPS,Robinson:2020:ICLR,Liu:2023:ICDE,Wang:2020:ACM,Qin:2020:AAAI,Chen:2021:ICLR}, the existing two backpropagation cannot fully address this issue as its loss gradient is still backpropagated to update user encodings. In contrast, our OneBP only applies his/her interacted items' encodings to update a user encoding, which can effectively decouple the impact of false negative samples.

\subsubsection{Simple but efficient}
Compared to the existing three categories of approaches for addressing the OCCF problem, our OneBP method offers a simpler and more efficient approach. Many neural encoding models may help to enhance recommendation performance, but they are often with complex structures and more parameters. Some negative sampling algorithms require additional computation for weighting unlabelled items and introduce many sensitive hyperparameters. Some other similarity and loss functions also need more computation time and space overheads. In contrast, our OneBP approach does not introduce complicated encoders and negative samplers, and its computation complexity is far smaller than others as the backpropagation is only for item encoding update. Our experiments in the next section also indicate better recommendation performance of our OneBP than that of peer algorithms.

\section{Experiments}\label{Sec:Experiments}

\subsection{Experiment Settings}

\textbf{Dataset}: We conduct experiments on four widely used public datasets, including the MovieLens-100k, MovieLens-1M~\footnote{MovieLens: https://grouplens.org/datasets/movielens/}, Gowalla~\footnote{Gowalla: https://paperswithcode.com/dataset/gowalla} and Yelp2018~\footnote{Yelp2018: https://www.yelp.com/dataset}. Table~\ref{tab:dataset} presents the statistics of the four datasets. We note that the Gowalla and Yelp2018 are often regarded as large datasets with serious interaction sparsity. In our research, we adopt a random selection process to create the training and test sets for each dataset. Specifically, we randomly select 80\% of the historical interactions for each user as the training set, while the remaining interactions are treated as the test set.


\begin{table}[t]
	\centering
	\caption{Dataset statistics}
	\renewcommand{\arraystretch}{0.4}
	\resizebox{0.4\textwidth}{!}
	{
		\begin{tabular}{lcccc}
			\toprule
			& \multicolumn{1}{l}{users} & \multicolumn{1}{l}{items}  & \multicolumn{1}{l}{interactions} & \multicolumn{1}{l}{density} \\
			\midrule
			MovieLens-100k & 943   & 1682    & 100k & 6.30\% \\
			MovieLens-1M & 6040  & 3952     & 1000k  & 4.19\% \\
			Gowalla & 29858 & 40981   & 1027k  & 0.08\% \\
			Yelp2018 & 31668 & 38048   & 1561k & 0.13\% \\
			\bottomrule
		\end{tabular}%
		\label{tab:dataset}%
	}
\end{table}%

\par
\textbf{Competitors}: We compare our proposed \textsf{OneBP} with the following three groups of competitors, as introduced in Section~\ref{Sec:RelatedWork}. (1) The first encoder group focuses on the design of user and item encoder, including the \textsf{MF}~\cite{Koren:2009:Computer}, \textsf{NGCF}~\cite{Wang:2019:SIGIR},  and \textsf{LightGCN}\cite{Xiangnan:2020:SIGIR}. (2) The second negative group focuses on negative sampling yet with the simple MF encoder, including the \textsf{PNS}~\cite{Mikolov:2013:NIPS,Chen:2017:KDD,Tang:2015:WWW}, \textsf{DNS}~\cite{Zhang:2013:SIGIR}, \textsf{HarS}~\cite{Hsieh:2017:WWW}, and \textsf{2stS}~\cite{Tran:2019:SIGIR}. (3) The third loss group focuses on the loss function, including the \textsf{CML}~\cite{Hsieh:2017:WWW}, \textsf{SFCML}~\cite{Bao:2022:IEEE}, \textsf{DPCML}~\cite{Bao:2022:NIPS}, \textsf{DCL}~\cite{Chuang:2020:NIPS},  \textsf{HCL}~\cite{Robinson:2020:ICLR}, and \textsf{BUIR}~\cite{Lee:2021:SIGIR}.

\par
\textbf{Evaluation metrics}: We adopt the mostly used performance metrics in the OCCF problem for personalized top-$K$ recommendation, including Precision (P), Recall (R), F1-Score (F1), and Normalized Discounted Cumulative Gain (NDCG). For their wide usage, we do not detail their calculations here. We use them to evaluate the top-$5$, top-$10$, and top-$20$ recommendation performance.


\textbf{Parameter Settings}:
To make fair comparisons, we use the same number of negative samples for all algorithms. More detailed hyperparameter settings are documented in the README file of the code. The codes is implemented with Pytorch~\footnote{The experimental codes of OneBP model in the paper are released at: ~\url{https://anonymous.4open.science/r/OneBP-7C87}}.

\subsection{Overall results}
Table~\ref{Tbl:OverallResult} compares the overall results of recommendation performance on the four datasets. We report that our \textsf{OneBP} outperforms all kinds of the state-of-the-art competitors, and the improvements over the second-best ones are with non-negligible margins. Recall that our \textsf{OneBP} adopts the simplest projection matrices as encoders, naive random negative sampling, and plain InfoNCE loss with dot product similarity. The remarkable performance of our \textsf{OneBP} should be only attributed to the effectiveness of the proposed one backpropagation strategy in model training.

\par
Table~\ref{Tbl:OverallResult} also compares the computation time per epoch on the four datasets, where our \textsf{OneBP} requires the least ones. The savings of computation time become more significant in larger datasets. This is not unexpected. As our solution does not require advanced neural encoding, sophisticated negative sampling and complicated loss computing, its computation complexity for model training can be greatly reduced. Further computation savings can be attributed to that our \textsf{OneBP} does not need to compute loss gradients for updating users' encodings.

\par
An illuminating comparison can be made between the \textsf{OneBP} and \textsf{MF}. The only difference lies in that the \textsf{MF} uses two gradient backpropagation to update encodings: one for items and another for users. The \textsf{MF} enjoys the second-best computation saver and our \textsf{OneBP} the best one, which suggests that cutting off user side gradient backpropagation can further reduce computing overhead. On the other hand, the recommendation performance of the \textsf{MF} are often worse than those state-of-the-art competitors, but our \textsf{OneBP} plays the best.

\par
The overall results confirm that the simple one backpropagation updating strategy in our \textsf{OneBP} has great advantages in terms of recommendation effectiveness and computation efficiency.

\begin{table*}[!]
	\centering
	\caption{Overall comparison of recommendation performance.}\label{Tbl:OverallResult}
	\renewcommand{\arraystretch}{0.9}
	\resizebox{\textwidth}{!}
	{
		\begin{threeparttable}
			\begin{tabular}{cclccccccccccccc}
				\toprule
				Dataset & Type  & Method & P@5   & R@5   & F1@5  & NDCG@5 & P@10  & R@10  & F1@10 & NDCG@10 & P@20  & R@20  & F1@20 & NDCG@20 & time(s) \\
				\midrule
				\multicolumn{1}{c}{\multirow{16}[8]{*}{MovieLens\newline{}-100k}} & \multicolumn{1}{r}{\multirow{3}[2]{*}{Encoder}}
				& MF   & 0.4155 & 0.1434 & 0.1850 & 0.4430 & 0.3496 & 0.2308 & 0.2328 & 0.4181 & 0.2824 & 0.3529 & 0.2606 & 0.4183 & \cellcolor{orange!30}1.19\\
				& & LightGCN & \cellcolor{orange!15}0.4263 & \cellcolor{orange!15}0.1458 & \cellcolor{orange!15}0.1881 & \cellcolor{orange!8}0.4531 & \cellcolor{orange!30}0.3587 & \cellcolor{orange!15}0.2313 & \cellcolor{orange!15}0.2351 & \cellcolor{orange!15}0.4269 & \cellcolor{orange!30}0.2877 & \cellcolor{orange!15}0.3474 & \cellcolor{orange!15}0.2609 & \cellcolor{orange!15}0.4236 & 2.17\\
				& & NGCF  & 0.4176 & 0.1452 & 0.1871 & 0.4469 & 0.3582 & 0.2356 & 0.238 & 0.4264 & 0.2875 & 0.3597 & 0.2652 & 0.4248 & 3.21 \\
				
				\cmidrule{2-16} & \multirow{4}[2]{*}{Negative}
				& DNS   & \cellcolor{orange!8}0.4214 & \cellcolor{orange!8}0.1460 & \cellcolor{orange!8}0.1882 & \cellcolor{orange!8}0.4501 & \cellcolor{orange!8}0.3531 & \cellcolor{orange!8}0.2317 & \cellcolor{orange!8}0.2342 & \cellcolor{orange!8}0.4226 & \cellcolor{orange!8}0.2836 & \cellcolor{orange!8}0.3544 & \cellcolor{orange!8}0.2614 & \cellcolor{orange!8}0.4228 & 1.83 \\
				& & PNS   & 0.3979 & 0.1338 & 0.1742 & 0.4231 & 0.3415 & 0.218 & 0.223 & 0.4026 & 0.2741 & 0.3373 & 0.2511 & 0.4016 & 12.09\\
				& & HarS  &\cellcolor{orange!8} 0.4227 & \cellcolor{orange!8}0.1455 &\cellcolor{orange!8} 0.1878 &\cellcolor{orange!8} 0.4516 &\cellcolor{orange!8} 0.3527 &\cellcolor{orange!8} 0.2312 &\cellcolor{orange!8} 0.2342 &\cellcolor{orange!8} 0.4239 &\cellcolor{orange!8} 0.2852 &\cellcolor{orange!8} 0.3557 & \cellcolor{orange!8}0.2624 &\cellcolor{orange!8} 0.4247 & 1.59\\
				& & 2stS  & 0.3718 & 0.1217 & 0.1597 & 0.3929 & 0.3188 & 0.2027 & 0.2085 & 0.3715 & 0.2574 & 0.3184 & 0.2365 & 0.3717 & 13.45 \\
				
				\cmidrule{2-16} & \multirow{6}[2]{*}{Loss} 
				& CML   & 0.3796 & 0.1174 & 0.1562 & 0.3963 & 0.3314 & 0.2022 & 0.2109 & 0.3793 & 0.272 & 0.3243 & 0.2458 & 0.3811 & 1.45\\
				& & SFCML & 0.4174 & 0.1432 & 0.1846 & 0.4429 & 0.3544 & 0.2292 & 0.2321 & 0.4202 & 0.2814 & 0.3389 & 0.2542 & 0.4147 & 2.61\\
				& & DPCML & 0.3869 & 0.1337 & 0.1717 & 0.4115 & 0.3207 & 0.2108 & 0.213 & 0.3836 & 0.2647 & 0.3287 & 0.2435 & 0.3833 & \cellcolor{orange!8}1.55\\
				& & DCL   & \cellcolor{orange!30}0.4291 & \cellcolor{orange!30}0.1501 & \cellcolor{orange!30}0.1928 &\cellcolor{orange!30} 0.4587 & \cellcolor{orange!30}0.3608 & \cellcolor{orange!30}0.2384 & \cellcolor{orange!30}0.2402 & \cellcolor{orange!30}0.4319 & \cellcolor{orange!30}0.2845 & \cellcolor{orange!30}0.3542 &\cellcolor{orange!30} 0.2621 & \cellcolor{orange!30}0.4257 & \cellcolor{orange!15}1.24\\
				& & HCL   & \cellcolor{orange!15}0.4263 & \cellcolor{orange!30}0.1499 & \cellcolor{orange!30}0.1924 &\cellcolor{orange!15} 0.4576 & \cellcolor{orange!15}0.3576 & \cellcolor{orange!30}0.2365 & \cellcolor{orange!30}0.2389 & \cellcolor{orange!30}0.4295 & \cellcolor{orange!15}0.2814 & \cellcolor{orange!15}0.3513 & \cellcolor{orange!15}0.2603 & \cellcolor{orange!15}0.4229 & \cellcolor{orange!15}1.26\\
				& & BUIR  & 0.4146 & 0.1387 & 0.1804 & 0.4373 & 0.3491 & 0.2258 & 0.2297 & 0.4121 & 0.2743 & 0.3445 & 0.2529 & 0.407 & \cellcolor{orange!8}1.57\\
				
				\cmidrule{2-16} & \multirow{2}[2]{*}{Proposed}
				&\multirow{2}[2]{*}{\textbf{OneBP}} & \cellcolor{orange!45}\textbf{0.453} & \cellcolor{orange!45}\textbf{0.1559} & \cellcolor{orange!45}\textbf{0.2015} & \cellcolor{orange!45}\textbf{0.4822} & \cellcolor{orange!45}\textbf{0.3811} & \cellcolor{orange!45}\textbf{0.2494} & \cellcolor{orange!45}\textbf{0.2529} & \cellcolor{orange!45}\textbf{0.4535} & \cellcolor{orange!45}\textbf{0.3032} & \cellcolor{orange!45}\textbf{0.373} & \cellcolor{orange!45}\textbf{0.2776} & \cellcolor{orange!45}\textbf{0.4501} & \cellcolor{orange!45}\textbf{1.11}\\
				& &       & \cellcolor{orange!45}5.57\%$\uparrow$ &\cellcolor{orange!45} 3.86\%$\uparrow$ &\cellcolor{orange!45} 4.51\%$\uparrow$ &\cellcolor{orange!45} 5.12\%$\uparrow$ &\cellcolor{orange!45} 5.63\%$\uparrow$ &\cellcolor{orange!45} 4.61\%$\uparrow$ &\cellcolor{orange!45} 5.29\%$\uparrow$ &\cellcolor{orange!45} 5.00\%$\uparrow$ &\cellcolor{orange!45} 5.39\%$\uparrow$ &\cellcolor{orange!45} 3.70\%$\uparrow$ &\cellcolor{orange!45} 4.68\%$\uparrow$ &\cellcolor{orange!45} 5.73\%$\uparrow$ &\cellcolor{orange!45} 6.72\%$\downarrow$\\
				\midrule
				\multicolumn{1}{c}{\multirow{16}[8]{*}{MovieLens\newline{}-1M}} & \multicolumn{1}{r}{\multirow{3}[2]{*}{Encoder}}
				& MF   & 0.4046 & 0.0962 & 0.1355 & 0.4234 & 0.3510 & 0.1584 & 0.1816 & 0.3930 & 0.2922 & 0.2477 & 0.2166 & 0.3780 & \cellcolor{green!25}14.28 \\
				& & LightGCN & 0.4121 & 0.0955 & 0.1350 & 0.4327 & 0.3537 & 0.1584 & 0.1792 & 0.3981 & 0.2921 & 0.2401 & 0.2132 & 0.3791 & 47.34\\
				& & NGCF  & \cellcolor{green!25}0.4192 & \cellcolor{green!25}0.1014 & \cellcolor{green!25}0.1422 & \cellcolor{green!25}0.4397 &\cellcolor{green!25} 0.3608 & \cellcolor{green!25}0.1643 & \cellcolor{green!25}0.1879 & \cellcolor{green!25}0.4059 &\cellcolor{green!25} 0.2983 & \cellcolor{green!25}0.2560 &\cellcolor{green!25} 0.2234 & \cellcolor{green!25}0.3888 
				& 62.35\\
				
				\cmidrule{2-16}          & \multirow{4}[2]{*}{Negative}
				& DNS   & 0.4080 & \cellcolor{green!18}0.0975 & \cellcolor{green!18}0.1372 & 0.4276 & \cellcolor{green!18}0.3558 & \cellcolor{green!18}0.1606 &\cellcolor{green!18} 0.1841 & \cellcolor{green!18}0.3980 & \cellcolor{green!18}0.2948 & \cellcolor{green!18}0.2504 & \cellcolor{green!18}0.2190 & \cellcolor{green!18}0.3819 & 19.36\\
				& & PNS   & 0.3931 & 0.0950 & 0.1334 & 0.4129 & 0.3362 & \cellcolor{green!8}0.1542 & 0.1759 & 0.3802 & 0.2773 & 0.2386 & 0.2079 & 0.3639 & 254.74\\
				& & HarS  & 0.3453 & 0.0719 & 0.1042 & 0.3687 & 0.2770 & 0.1086 & 0.1293 & 0.3201 & 0.2049 & 0.1544 & 0.1424 & 0.2781 & 18.24 \\
				& & 2stS  & 0.3772 & 0.0930 & 0.1295 & 0.3978 & 0.3261 & 0.1507 & 0.1704 & 0.3679 & 0.2689 & 0.2362 & 0.2029 & 0.3521 & 276.35 \\
				
				\cmidrule{2-16} & \multirow{6}[2]{*}{Loss} 
				& CML   & 0.3425 & 0.0753 & 0.1077 & 0.3554 & 0.3047 & 0.1277 & 0.1497 & 0.3341 & 0.258 & 0.2055 & 0.1847 & 0.3226 & \cellcolor{green!18}16.39\\
				& & SFCML & 0.3968 & \cellcolor{green!8}0.0959 & \cellcolor{green!8}0.1348 & 0.4136 & \cellcolor{green!8}0.3481 & \cellcolor{green!18}0.1573 & \cellcolor{green!18}0.1808 & \cellcolor{green!8}0.3875 & \cellcolor{green!18}0.2912 & \cellcolor{green!18}0.243 & \cellcolor{green!8}0.2154 & \cellcolor{green!18}0.3741 & 57.02\\
				& & DPCML & 0.3883 & 0.088 & 0.125 & 0.4079 & 0.3359 & 0.1424 & 0.1665 & 0.3755 & 0.2758 & 0.2198 & 0.1983 & 0.3538 & 16.89\\
				& & DCL   & \cellcolor{green!18}0.4133 & 0.0953 & \cellcolor{green!8}0.1350 &\cellcolor{green!18} \cellcolor{green!18}0.4355 & \cellcolor{green!18}0.3512 &\cellcolor{green!8} 0.152 & \cellcolor{green!8}0.177 & \cellcolor{green!18}0.3967 & \cellcolor{green!8}0.2867 &\cellcolor{green!8} 0.2328 & \cellcolor{green!8}0.2084 & \cellcolor{green!8}0.3732 & \cellcolor{green!25}14.24 \\
				& & HCL   & \cellcolor{green!8}0.4125 & \cellcolor{green!8}0.0957 & \cellcolor{green!8}0.1355 & \cellcolor{green!8}0.4345 & \cellcolor{green!18}0.3515 &\cellcolor{green!8} 0.1528 & \cellcolor{green!8}0.1776 & \cellcolor{green!18}0.3972 & \cellcolor{green!8}0.2867 & \cellcolor{green!8}0.2336 & \cellcolor{green!8}0.2086 & \cellcolor{green!8}0.3736 & \cellcolor{green!8}14.62\\
				& & BUIR  & 0.3875 & 0.0942 & 0.1318 & 0.4089 & 0.3314 & \cellcolor{green!8}0.152 & 0.1728 & 0.376 & 0.2714 &\cellcolor{green!8} 0.2332 & 0.2023 & 0.3581 & 16.58 \\
				
				\cmidrule{2-16} & \multirow{2}[2]{*}{Proposed} 
				& \multirow{2}[2]{*}{\textbf{OneBP}} &\cellcolor{green!35} \textbf{0.4300} &\cellcolor{green!35} \textbf{0.1071} &\cellcolor{green!35} \textbf{0.1487} &\cellcolor{green!35} \textbf{0.4512} &\cellcolor{green!35} \textbf{0.3709} &\cellcolor{green!35} \textbf{0.1746} &\cellcolor{green!35} \textbf{0.1966} &\cellcolor{green!35} \textbf{0.4182} &\cellcolor{green!35} \textbf{0.3069} &\cellcolor{green!35} \textbf{0.2690} &\cellcolor{green!35} \textbf{0.2316} &\cellcolor{green!35} \textbf{0.4036} & \cellcolor{green!35}\textbf{14.21}\\
				& &       &\cellcolor{green!35} 2.58\%$\uparrow$ &\cellcolor{green!35} 5.62\%$\uparrow$ &\cellcolor{green!35} 4.57\%$\uparrow$ &\cellcolor{green!35} 2.62\%$\uparrow$ &\cellcolor{green!35} 2.80\%$\uparrow$ &\cellcolor{green!35} 6.27\%$\uparrow$ &\cellcolor{green!35} 4.63\%$\uparrow$ &\cellcolor{green!35} 3.03\%$\uparrow$ &\cellcolor{green!35} 2.88\%$\uparrow$ &\cellcolor{green!35} 5.08\%$\uparrow$ &\cellcolor{green!35} 3.67\%$\uparrow$ &\cellcolor{green!35} 3.81\%$\uparrow$ &\cellcolor{green!35} 0.49\%$\downarrow$ \\
				\midrule
				\multirow{16}[8]{*}{Gowalla} & \multicolumn{1}{r}{\multirow{3}[2]{*}{Encoder}}
				& MF   & 0.0739 & 0.0757 & 0.0621 & 0.1016 & 0.0560 &  0.1122 & 0.0623 & 0.1076 & 0.0422 & 0.1650 & 0.0587 & 0.1230 & \cellcolor{blue!20}20.91 \\
				& & LightGCN & \cellcolor{blue!9}0.0845 & \cellcolor{blue!14}0.0888 & \cellcolor{blue!9}0.0725 & \cellcolor{blue!9}0.1166 & \cellcolor{blue!9}0.0631 & \cellcolor{blue!9}0.1297 & \cellcolor{blue!9}0.0713 & \cellcolor{blue!9}0.1232 & \cellcolor{blue!9}0.0465 & \cellcolor{blue!9}0.1864 & \cellcolor{blue!9}0.0642 & \cellcolor{blue!9}0.1394 & 273.5\\
				& & NGCF  & 0.0768 & 0.0808 & 0.0661 & 0.1051 & 0.0585 & 0.121 & 0.0664 & 0.1127 & 0.044 & 0.1769 & 0.0608 & 0.1291 & 157.74\\
				
				\cmidrule{2-16}          & \multirow{4}[2]{*}{Negative}
				& DNS   & 0.0737 & 0.0779 & 0.0633 & 0.1013 & 0.0563 & 0.1162 & 0.0636 & 0.1085 & 0.0421 & 0.1696 & 0.0580 & 0.1240 & 37.05 \\
				& & PNS   & 0.0586 & 0.0611 & 0.0502 & 0.0811 & 0.0439 & 0.0894 & 0.0494 & 0.0855 & 0.0323 & 0.1298 & 0.0445 & 0.0968 & 3728.37  \\
				& & HarS  & 0.0719 & 0.0742 & 0.0609 & 0.0989 & 0.0544 & 0.1098 & 0.0607 & 0.1048 & 0.0411 & 0.16  & 0.0558 & 0.1189 & \cellcolor{blue!9}31.52 \\
				& & 2stS  & 0.0609 & 0.0599 & 0.0501 & 0.0832 & 0.0461 & 0.0883 & 0.0502 & 0.0871 & 0.0345 & 0.1285 & 0.0462 & 0.0982 & 3125.77\\
				
				\cmidrule{2-16} & \multirow{6}[2]{*}{Loss} 
				& CML   & 0.0501 & 0.0603 & 0.0472 & 0.0705 & 0.0384 & 0.0901 & 0.0466 & 0.0782 & 0.0284 & 0.1285 & 0.0412 & 0.0902 & 91.11 \\
				& & SFCML & \cellcolor{blue!20}0.0899 & \cellcolor{blue!20}0.0931 & \cellcolor{blue!20}0.0766 & \cellcolor{blue!20}0.1243 & \cellcolor{blue!20}0.0672 & \cellcolor{blue!20}0.1365 & \cellcolor{blue!20}0.0756 & \cellcolor{blue!20}0.1278 &\cellcolor{blue!20} 0.0496 & \cellcolor{blue!20}0.1963 & \cellcolor{blue!20}0.0689 & \cellcolor{blue!20}0.1477 & 1842.02\\
				& & DPCML & 0.0782 & 0.0759 & 0.0642 & 0.1042 & 0.059 & 0.1124 & 0.0645 & 0.1092 & 0.0432 & 0.1632 & 0.0584 & 0.1225 & 92.13\\
				& & DCL   & 0.0761 & 0.0776 & 0.0642 & 0.1036 & 0.0575 & 0.1149 & 0.064 & 0.1093 & 0.0424 & 0.1651 & 0.0578 & 0.1236 & \cellcolor{blue!14}21.12\\
				& & HCL   & \cellcolor{blue!14}0.0873 & \cellcolor{blue!9}0.0885 & \cellcolor{blue!14}0.0731 & \cellcolor{blue!14}0.118 & \cellcolor{blue!14}0.0653 & \cellcolor{blue!9}0.1286 & \cellcolor{blue!14}0.0723 & \cellcolor{blue!14}0.1236 &\cellcolor{blue!9}0.0477 & \cellcolor{blue!9}0.1846 & \cellcolor{blue!9}0.0649 & \cellcolor{blue!9}0.1392 & \cellcolor{blue!20}20.98\\
				& & BUIR  & \cellcolor{blue!9}0.0854 & \cellcolor{blue!9}0.0872 &\cellcolor{blue!9} 0.072 &\cellcolor{blue!9}0.1166 &\cellcolor{blue!14} 0.0651 & \cellcolor{blue!14}0.1306 & \cellcolor{blue!14}0.0725 & \cellcolor{blue!14}0.1241 & \cellcolor{blue!14}0.0484 &\cellcolor{blue!14}0.1898 &\cellcolor{blue!14} 0.0661 & \cellcolor{blue!14}0.1411 & 30.17\\
				
				\cmidrule{2-16} & \multirow{2}[2]{*}{Proposed} 
				& \multirow{2}[2]{*}{\textbf{OneBP}} & \cellcolor{blue!30}\textbf{0.0948} & \cellcolor{blue!30}\textbf{0.0971} & \cellcolor{blue!30}\textbf{0.0803} & \cellcolor{blue!30}\textbf{0.1282} & \cellcolor{blue!30}\textbf{0.0715} & \cellcolor{blue!30}\textbf{0.1431} & \cellcolor{blue!30}\textbf{0.0798} & \cellcolor{blue!30}\textbf{0.1356} & \cellcolor{blue!30}\textbf{0.0527} & \cellcolor{blue!30}\textbf{0.2059} & \cellcolor{blue!30}\textbf{0.0722} & \cellcolor{blue!30}\textbf{0.1536} & \cellcolor{blue!30}\textbf{20.12} \\
				& &       & \cellcolor{blue!30}5.45\%$\uparrow$ & \cellcolor{blue!30}4.30\%$\uparrow$ & \cellcolor{blue!30}4.83\%$\uparrow$ & \cellcolor{blue!30}3.14\%$\uparrow$ & \cellcolor{blue!30}6.40\%$\uparrow$ & \cellcolor{blue!30}4.84\%$\uparrow$ & \cellcolor{blue!30}5.56\%$\uparrow$ & \cellcolor{blue!30}6.10\%$\uparrow$ & \cellcolor{blue!30}6.25\%$\uparrow$ & \cellcolor{blue!30}4.89\%$\uparrow$ & \cellcolor{blue!30}4.79\%$\uparrow$ & \cellcolor{blue!30}3.99\%$\uparrow$ &\cellcolor{blue!30} 3.78\%$\downarrow$ \\
				\midrule
				\multirow{16}[8]{*}{Yelp2018} & \multicolumn{1}{r}{\multirow{3}[2]{*}{Encoder}}
				& MF   & 0.0429 & 0.0246 & 0.0284 & 0.0470 &  0.0365 & 0.0417 & 0.0352 & 0.0491 & 0.0305 & 0.0700 & 0.0394 &  0.0580 & \cellcolor{yellow!40}44.02 \\
				& & LightGCN & \cellcolor{yellow!20}0.0527 & \cellcolor{yellow!20}0.0318 & \cellcolor{yellow!20}0.0354 & \cellcolor{yellow!20}0.0577 & \cellcolor{yellow!20}0.0452 & \cellcolor{yellow!20}0.0536 & \cellcolor{yellow!20}0.0430 & \cellcolor{yellow!20}0.0613 & \cellcolor{yellow!20}0.0374 & \cellcolor{yellow!20}0.0884 & \cellcolor{yellow!20}0.0464 & \cellcolor{yellow!20}0.0724 & 1155.24\\
				& & NGCF  & 0.0511 & 0.0307 & 0.0343 & 0.0555 & 0.0438 & 0.0523 & 0.0419 & 0.059 & 0.0364 & 0.0865 & 0.0453 & 0.0702 & 502.13\\	
				\cmidrule{2-16}          & \multirow{4}[2]{*}{Negative}
				& DNS   & 0.0469 & 0.0279 & 0.0312 & 0.0509 & 0.0404 & 0.0481 & 0.0385 & 0.0541 & 0.0341 & 0.0804 & 0.0422 & 0.0647 & 73.58\\
				& & PNS   & 0.0324 & 0.0196 & 0.0218 & 0.0351 & 0.0280 & 0.0335 & 0.0267 & 0.0374 & 0.0235 & 0.0555 & 0.0291 & 0.0448 & 5526.33\\
				& & HarS  & 0.0418 & 0.0241 & 0.0273 & 0.0456 & 0.0353 & 0.0407 & 0.0331 & 0.0474 & 0.0295 & 0.0678 & 0.0362 & 0.0559 & \cellcolor{yellow!20}54.75\\
				& & 2stS  & 0.0393 & 0.0234 & 0.0261 & 0.0429 & 0.0335 & 0.0397 & 0.0319 & 0.0454 & 0.0277 & 0.0654 & 0.0342 & 0.0536 & 4521.21\\
				
				\cmidrule{2-16} & \multirow{6}[2]{*}{Loss} 
				& CML   & 0.0347 & 0.0221 & 0.0245 & 0.0397 & 0.0309 & 0.0386 & 0.0301 & 0.0398 & 0.0261 & 0.0658 & 0.0314 & 0.0501 & 143.08\\
				& & SFCML & \cellcolor{yellow!30}0.0552 & \cellcolor{yellow!30}0.0332 & \cellcolor{yellow!30}0.0370 & \cellcolor{yellow!30}0.0605 & \cellcolor{yellow!30}0.0469 & \cellcolor{yellow!30}0.0556 & \cellcolor{yellow!30}0.0448 & \cellcolor{yellow!30}0.0637 &\cellcolor{yellow!30} 0.0384 &\cellcolor{yellow!30} 0.0903 & \cellcolor{yellow!30}0.0477 & \cellcolor{yellow!30}0.0747 & 3482.52\\
				& & DPCML & 0.0467 & 0.0273 & 0.0307 & 0.0506 & 0.0393 & 0.0459 & 0.0371 & 0.053 & 0.0328 & 0.076 & 0.0403 & 0.0626 & 144.7\\
				& & DCL   & 0.0485 & 0.0285 & 0.0321 & 0.0528 & 0.0415 & 0.0485 & 0.0391 & 0.0557 & 0.0346 & 0.0804 & 0.0426 & 0.0656 & \cellcolor{yellow!30}44.29\\
				& & HCL   & \cellcolor{yellow!20}0.054 & \cellcolor{yellow!20}0.0327 & \cellcolor{yellow!20}0.0365 & \cellcolor{yellow!20}0.0591 & \cellcolor{yellow!20}0.0456 & \cellcolor{yellow!20}0.0544 & \cellcolor{yellow!20}0.0436 & \cellcolor{yellow!20}0.0621 & \cellcolor{yellow!20}0.0377 & \cellcolor{yellow!20}0.0886 & \cellcolor{yellow!20}0.0464 & \cellcolor{yellow!20}0.0726 & \cellcolor{yellow!30}44.5\\
				& & BUIR  & \cellcolor{yellow!40}0.0597 & \cellcolor{yellow!40}0.0364 & \cellcolor{yellow!40}0.0404 & \cellcolor{yellow!40}0.0655 &\cellcolor{yellow!40} 0.0501 &\cellcolor{yellow!40} 0.0603 &\cellcolor{yellow!40} 0.0481 &\cellcolor{yellow!40} 0.0688 &\cellcolor{yellow!40} 0.041 &\cellcolor{yellow!40} 0.0982 &\cellcolor{yellow!40} 0.0512 &\cellcolor{yellow!40} 0.0809 & 64.8\\
				
				\cmidrule{2-16} & \multirow{2}[2]{*}{Proposed} 
				& \multirow{2}[2]{*}{\textbf{OneBP}} & \cellcolor{yellow!50}\textbf{0.0616} & \cellcolor{yellow!50}\textbf{0.0366} & \cellcolor{yellow!50}\textbf{0.041} & \cellcolor{yellow!50}\textbf{0.0675} & \cellcolor{yellow!50}\textbf{0.0521} & \cellcolor{yellow!50}\textbf{0.0613} & \cellcolor{yellow!50}\textbf{0.0494} & \cellcolor{yellow!50}\textbf{0.0708} & \cellcolor{yellow!50}\textbf{0.0428} & \cellcolor{yellow!50}\textbf{0.0997} & \cellcolor{yellow!50}\textbf{0.0527} & \cellcolor{yellow!50}\textbf{0.0827} & \cellcolor{yellow!50}\textbf{36.33}\\
				& &       & \cellcolor{yellow!50}3.18\%$\uparrow$ & \cellcolor{yellow!50}0.55\%$\uparrow$ & \cellcolor{yellow!50}1.49\%$\uparrow$ & \cellcolor{yellow!50}3.05\%$\uparrow$ & \cellcolor{yellow!50}3.99\%$\uparrow$ & \cellcolor{yellow!50}1.66\%$\uparrow$ & \cellcolor{yellow!50}2.70\%$\uparrow$ & \cellcolor{yellow!50}2.91\%$\uparrow$ & \cellcolor{yellow!50}4.39\%$\uparrow$ & \cellcolor{yellow!50}1.53\%$\uparrow$ & \cellcolor{yellow!50}2.93\%$\uparrow$ & \cellcolor{yellow!50}2.22\%$\uparrow$ & \cellcolor{yellow!50}17.5\%$\downarrow$\\
				\bottomrule
			\end{tabular}%
			
			\begin{tablenotes}
				\footnotesize
				\item[1] The value ahead $\uparrow$ and $\downarrow$ measures the improvement and degradation percentage compared to the optimal value among other baselines.
				\item[2] The color represents the degree of excellence in recommending performance. The darker the color, the better the corresponding metric.
			\end{tablenotes}
		\end{threeparttable}
	}
\end{table*}%

	%

\subsection{OneBP on different encoders}
We note that the encoding update operation in our \textsf{OneBP} can also be applied to other encoders. This can be easily done by first updating the inputs of a user/item encoder according to our OneBP encoding update operation and next using the updated inputs into the user/item encoder. Following this idea, we experiments on the \textsf{NGCF} and \textsf{LightGCN} encoder by implementing our updating operation as a preprocessing block before the original encoders.

\par
\begin{figure}[t]
	\centering
	\includegraphics[width=0.8\textwidth,height=0.4\textheight]{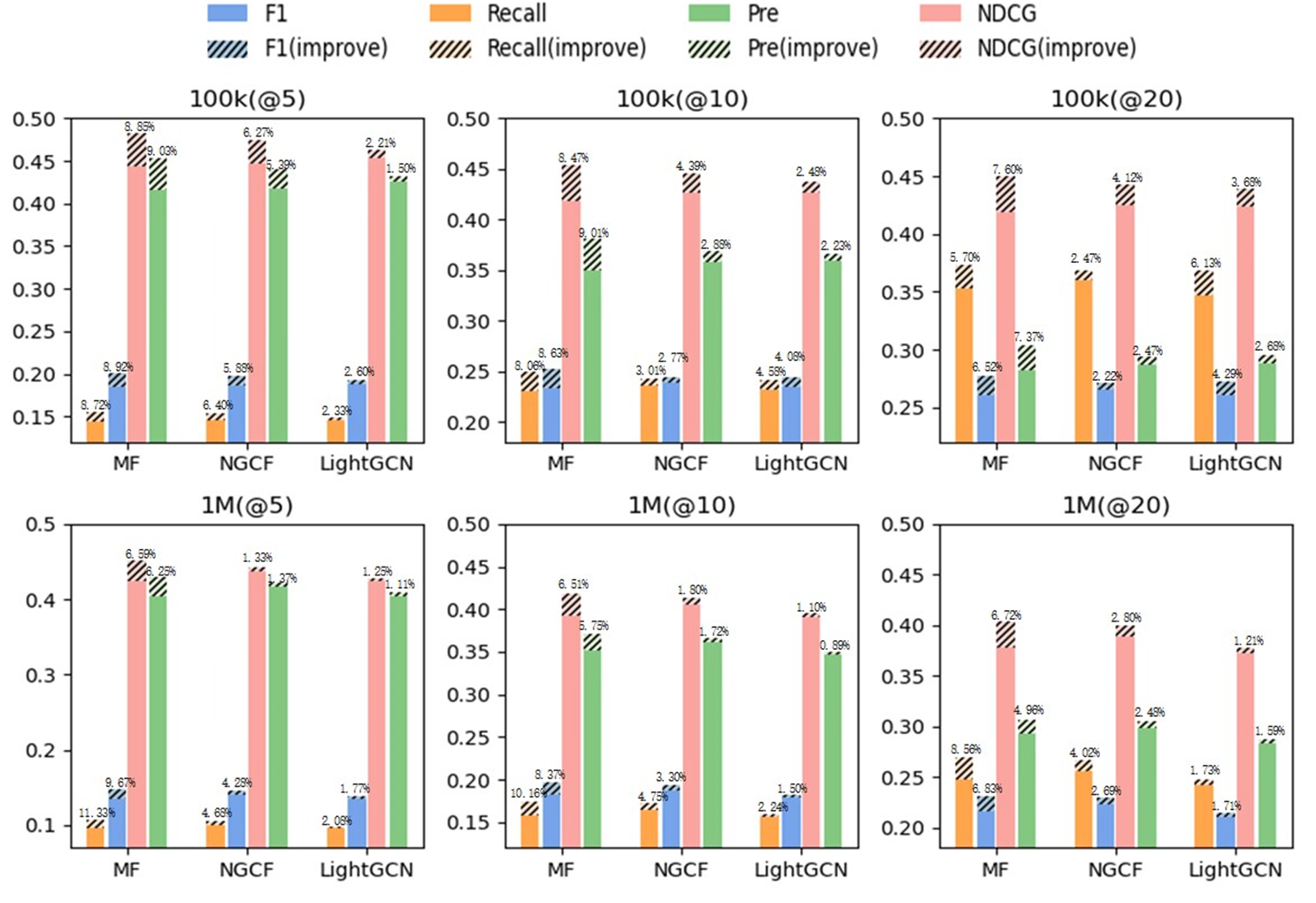}
	\caption{Performance of OneBP on different encoders.}	
	\label{Fig:OneBPOtherEncoder}
\end{figure}

Fig.~\ref{Fig:OneBPOtherEncoder} compares the recommendation performance for using original encoders and for using original encoders plus our \textsf{OneBP} updating operation. It is observed that our encoding update operation as a preprocessing block for original encoders can further help improving their performance. This indicates many potentials of applying the one backpropagation and update operation in recommender systems. We also note our \textsf{OneBP} can even outperform the \textsf{NGCF} and \textsf{LightGCN} plus our updating operation. This observation suggests that a good training strategy might be more important for two tower models to solve the OCCF problem.

\subsection{Hyper-parameter Study}
Our \textsf{OneBP} contains only one hyperparameter $\beta$ to balance the weight between a user encoding in last iteration and his/her interacted item encoding updated in this iteration, c.f., Eq.~\eqref{Eq:OneBPUser}. Note that the number of negatives $N_s$ is a common hyperparameter in contrastive learning. We conduct experiments to analyze the effects of the two hyperparameters.

\begin{figure*}[h]
	\centering
	\includegraphics[width=0.85\textwidth,height=0.23\textheight]{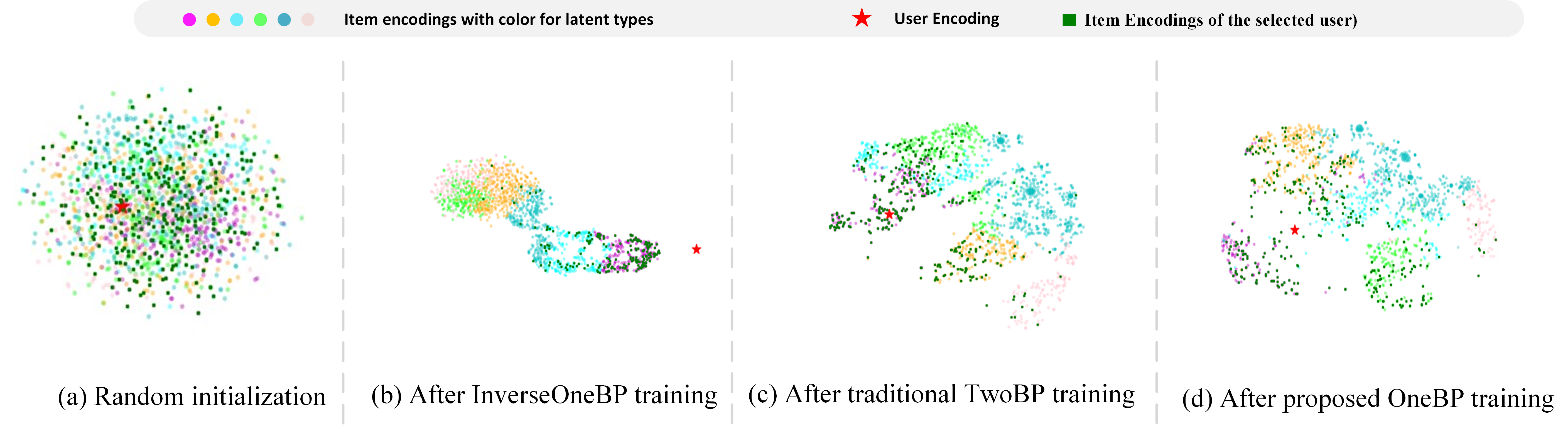}
	\caption{Visualization of clustering on items' encodings at initialization and after model training by different strategies.}
	\label{Fig:EncodingClustering}
\end{figure*}

\par
\begin{figure}[t]
	\centering
	\includegraphics[width=0.8\textwidth,height=0.38\textheight]{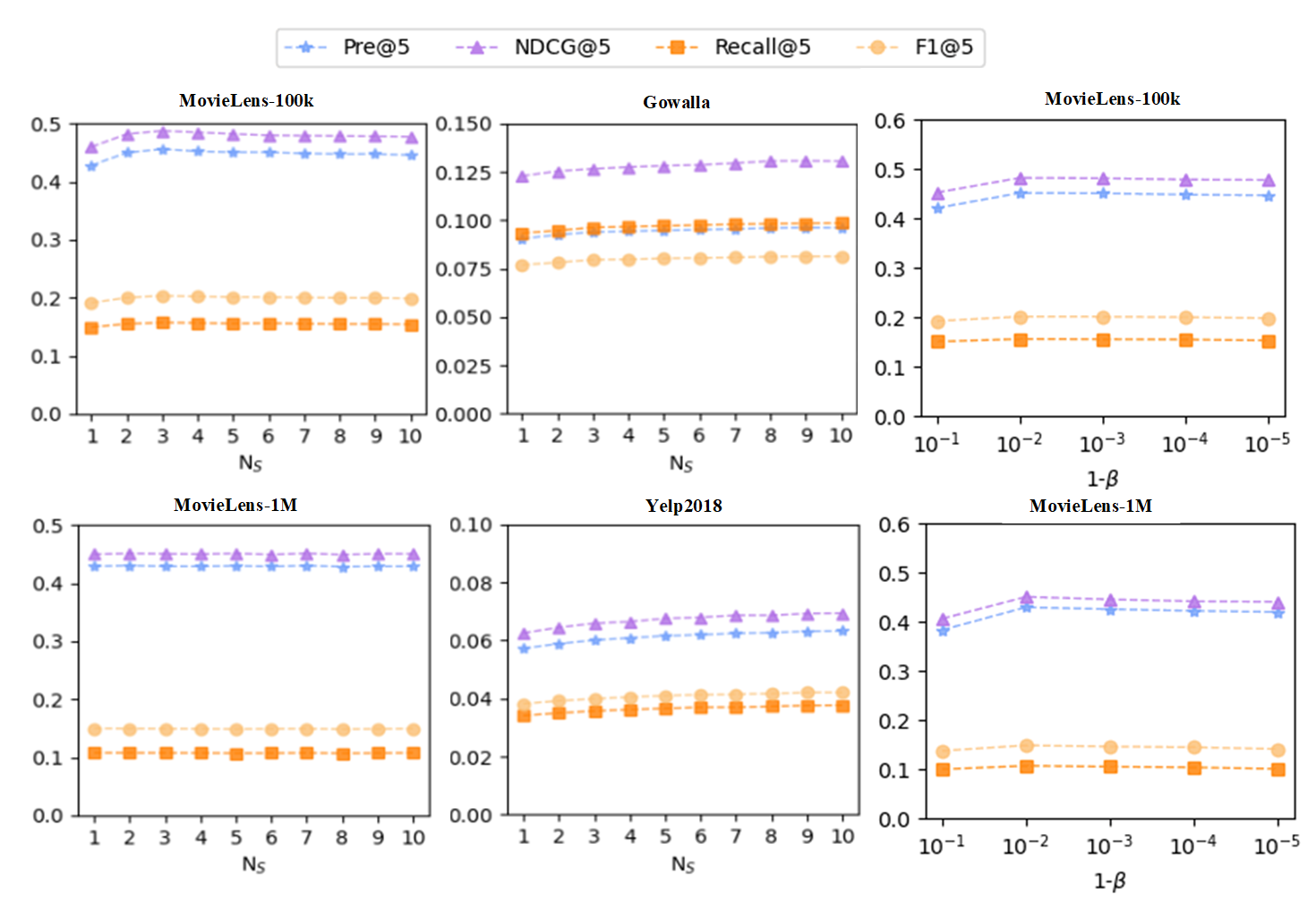}
	\caption{Hyper-parameter Study for $N_s$ and $\beta$}
	\label{Fig:Ns}
\end{figure}

\textbf{The hyperparameter} $\beta$: We vary the value of $1-\beta$ from $10^{-1}$ to $10^{-6}$ to analyze the recommendation sensitivity . Fig.~\ref{Fig:Ns} plots the performance metrics against the choice of this hyperparamter. It is observed that the recommendation performance slightly increases for $1-\beta$ decreasing from $10^{-1}$ to $10^{-2}$ and remains almost stable for $1-\beta<10^{-2}$. This first indicates that our \textsf{OneBP} is not much sensitive to the choice of $\beta$. Yet a small value of $1-\beta$ ($10^{-2}$ in the experiments) is suggested for the moving-aggregation updating strategy: A large value of $1-\beta$ means to pay more attention on those training items in the current batch while overlooking previous ones for user encoding update, leading to some undesirable variations.

\par
\textbf{The hyperparameter} $N_s$: It is well-known that the number of negatives impacts on the performance of contrastive learning. Generally, a larger number $N_s$, the better contrastive learning. However, this normally is also with the cost of increased computation complexity, and the adverse impacts of false negatives could become serious for more unlabeled data been selected as negatives. Fig.~\ref{Fig:Ns} plots the recommendation performance against the choices of negative numbers. Notice that the case of using $N_1=1$ in InfoNCE corresponds to the BPR loss. It is seen that using more negatives does help to improve recommendation performance. It is also observed that the improvements do not become significant for too more negatives on small scale dataset and we use $N_s=5$ in the comparision experiments.

\subsection{Exploration Study}\label{Sec:ExplorationStudy}
Recall that our \textsf{OneBP} uses gradient backpropagation only for items' encoding update. An interesting question would be that how about we use gradient backpropagation only for users' encoding update? That is, if we flip the update operation in Eqs.~\eqref{Eq:OneBPItem} and \eqref{Eq:OneBPUser}, how good is a two tower recommendation model?

\par
Table~\ref{Tbl:FlipOneBP} compares the recommendation performance by using the three backpropagation strategies, namely, both users and items, only users, and only items. Except using different backpropagation strategy, they all use the projection matrix-based user-item encoding, random negative sampling and plain InfoNCE loss. Results show that using one backpropagation for only item encoding update (our \textsf{OneBP}) outperforms using two backpropagation (traditional \textsf{TwoBP}) and using one backpropagation for only user encoding update. The results suggest that users' interests could be better learned from interacted items, but not the case that items types could be better learned from interacted users. As we argued before, with or without users' interactions, items are with intrinsic attributes and latent types. Note that in using one backpropagation for only user encoding update, items' encodings are updated based on the users' encodings, which may not well respect items' intrinsic attributes for item encoding.

\begin{table}[h]
	\centering
	\caption{Exploration study on backpropagation strategy.}
	\renewcommand{\arraystretch}{0.1}
	\resizebox{0.8\textwidth}{!}
	{
		\begin{tabular}{c|cc|llllll}
			\toprule
			Dataset & User  & Item  & P@5   & R@5   & P@10  & R@10  & P@20  & R@20 \\
			\midrule
			\multirow{3}[2]{*}{MovieLens-100k} & $\checkmark$     & $\checkmark$     & 0.4155 & 0.1434 & 0.3496 & 0.2308 & 0.2824 & 0.3529 \\
			& $\checkmark$     &       & 0.1088 & 0.0222 & 0.109 & 0.0411 & 0.1029 & 0.0777 \\
			&       & $\checkmark$     & \textbf{0.453} & \textbf{0.1559} & \textbf{0.3811} & \textbf{0.2494} & \textbf{0.3032} & \textbf{0.373} \\
			\midrule
			\multirow{3}[2]{*}{MovieLens-1M} & $\checkmark$     & $\checkmark$     & 0.4046 & 0.0962 & 0.351 & 0.1584 & 0.2922 & 0.2477 \\
			& $\checkmark$     &       & 0.1081 & 0.016 & 0.103 & 0.0312 & 0.0981 & 0.0595 \\
			&       & $\checkmark$     & \textbf{0.43}  & \textbf{0.1071} & \textbf{0.3709} & \textbf{0.1746} & \textbf{0.3069} & \textbf{0.269} \\
			\midrule
			\multirow{3}[2]{*}{Gowalla} & $\checkmark$     & $\checkmark$     & 0.0739 & 0.0757 & 0.056 & 0.1122 & 0.0422 & 0.165 \\
			& $\checkmark$     &       & 0.0761 & 0.0727 & 0.0612 & 0.117 & 0.0474 & 0.1775 \\
			&       & $\checkmark$     & \textbf{0.0948} & \textbf{0.0971} & \textbf{0.0715} & \textbf{0.1431} & \textbf{0.0527} & \textbf{0.2059} \\
			\midrule
			\multirow{3}[2]{*}{Yelp2018} & $\checkmark$     & $\checkmark$     & 0.0429 & 0.0246 & 0.0365 & 0.0417 & 0.0305 & 0.07 \\
			& $\checkmark$     &       & 0.0373 & 0.0209 & 0.034 & 0.0383 & 0.0304 & 0.0686 \\
			&       & $\checkmark$     & \textbf{0.0616} & \textbf{0.0366} & \textbf{0.0521} & \textbf{0.0613} & \textbf{0.0428} & \textbf{0.0997} \\
			\bottomrule
		\end{tabular}%
		\label{Tbl:FlipOneBP}%
	}
\end{table}%

\begin{figure*}[h]
	\centering
	\includegraphics[width=\textwidth]{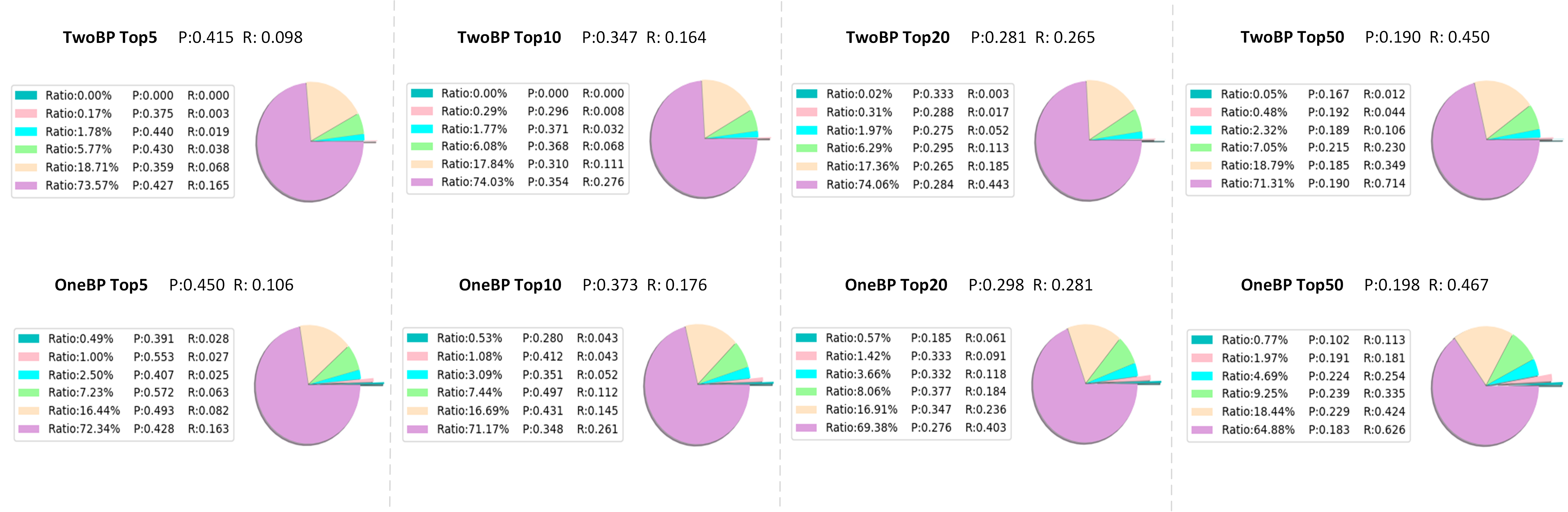}
	\caption{Recommendation statistics over all users when producing recommendation list with different lengths.}
	\label{Fig:PieAllUser}
\end{figure*}

\par

\begin{figure}[t]
	\centering
	\includegraphics[width=0.6\textwidth]{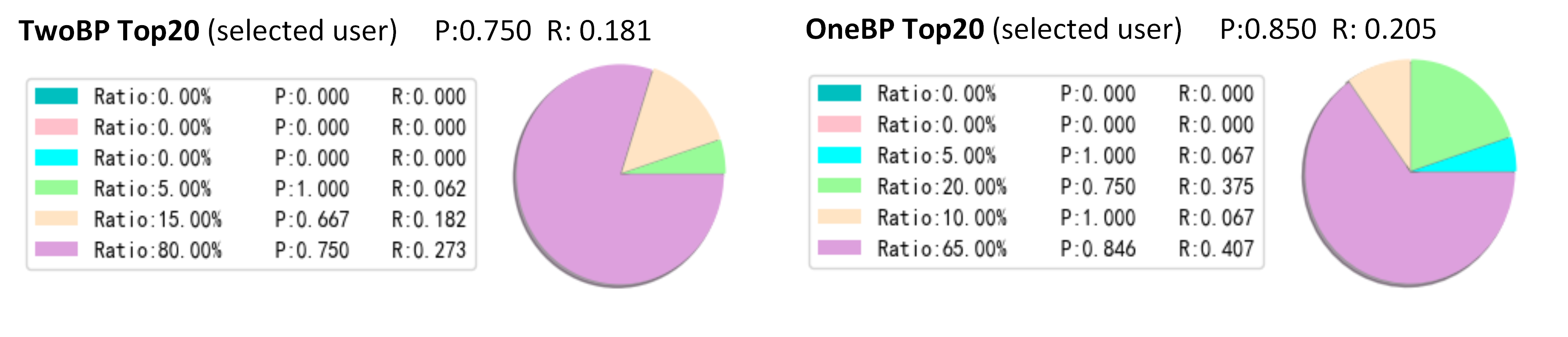}
	\caption{Recommendation statistics of the selected user.}
	\label{Fig:PieOneUser}
\end{figure}

Immediately following this, we explore how well OneBP captures user interest. However, data often lack labels in the OCCF problem, making it impossible to measure the quality of learned encodings as well as the diversity of user preferences. For items without labels, we resort to the KMeans clustering algorithm~\cite{Neyman:1967:Berkeley} to cluster items' encodings into six clusters, where one item belongs to one cluster. We use the t-SNE dimensionality reduction~\cite{Laurens:2008:JMLR} to map items' encodings into a two-dimensional plane, where colors indicate cluster indexes, i.e., latent types.

\par
Fig.~\ref{Fig:EncodingClustering} plots the distribution of projected items' encodings for both the training and testing dataset, where a user encoding is also projected and his/her interacted items in the training dataset are marked as rectangles and with dark green color. Fig.~\ref{Fig:EncodingClustering} (a) presents the distribution of items' encodings at the random initialization phase. We can observe that items with different latent types are almost mixed and there are no clear cluster divisions. Fig.~\ref{Fig:EncodingClustering} (b)(c)(d) present the distribution of items' encodings after the training of a two tower model. It can be clearly seen that items of the same color (indicating of the same latent type) are closer to each others as a cluster, and the divisions of different clusters are clearly observable. This validates our arguments to some extent, that is, a two tower recommendation model is capable of learning items' intrinsic yet unexposed attributes as encoded in the learned items' encodings, and items can be clustered into laten types with observable cluster divisions.

\par
Fig.~\ref{Fig:EncodingClustering} (b) displays the clustering results by using one backpropagation only for the user encoding update. It can be observed that the learned items' encodings do not present a kind of uniform dispersion in the space, and clusters' divisions are also not that clear, as compared with Fig.~\ref{Fig:EncodingClustering} (c) and (d).  Furthermore, the interacted items by a user in the training dataset are too much concentrated in one cluster, and the user encoding is farther away from his/her interacted items, also compared with Fig.~\ref{Fig:EncodingClustering} (c) and (d). These observations indicate that this user backpropagation training strategy may not be capable of learning high quality representations for items and for users, so its recommendation performance suffers.

\par
We next take a close comparison on the traditional \textsf{TwoBP} and our \textsf{OneBP} updating strategy. From Fig.~\ref{Fig:EncodingClustering} (c) and (d), we can observe that the interacted items by a user in the training dataset are largely distributed in one cluster, so-called majority items; While few are distributed in other clusters, so-called minority items, as discussed in Section~\ref{Sec:Symmetric}. Comparing the user representation learned by the \textsf{TwoBP} in Fig.~\ref{Fig:EncodingClustering} (c) and that by the \textsf{OneBP} in Fig.~\ref{Fig:EncodingClustering} (d), we can observe that the former is surrounded by majority items, yet being farther away from minority items; While the later is kind of located in the middle of majority items and minority items. The former indicates that the user representation learned by the \textsf{TwoBP} is kind of dominated by majority items, lack of some diversity. The later suggests that our \textsf{OneBP} is good at learning diverse interests for a user representation from both majority items and minority items.

\par
We next explore the recommendation effects of learning diverse interests for users. Fig.~\ref{Fig:PieOneUser} reports the Precision (P) and Recall (R) statistics in each latent type for the selected user in Fig.~\ref{Fig:EncodingClustering}. The percentage means the ratio of items of the same type over the length of the recommendation list. The left is by using the traditional \textsf{TwoBP}, where three latent types are identified for his/her testing items; The right is by using our \textsf{OneBP}, where four latent types are identified. Comparing the Precision and Recall for majority testing items, our \textsf{OneBP} outperforms the \textsf{TwoBP}. For latent types of minority testing items, our \textsf{OneBP} also performs better, so achieving better overall recommendation results. Fig.~\ref{Fig:PieAllUser} reports the top-$K$ recommendation statistics over all users. With the increase of $K$, our \textsf{OneBP} presents a clearer tendency of much reduced majority items. Furthermore, minority items are distributed more proportionally in different clusters. This suggests a better diversification for items. While such diversification gets paid in terms better recommendation performance for minority items in each latent type. As such, our \textsf{OneBP} achieves better overall performance than the \textsf{TwoBP}.

\section{Conclusion}\label{Sec:Conclusion}
This paper has proposed a new gradient backpropagation strategy for two tower recommendation models, called OneBP. The basic idea is to replacing the gradient backpropagation for the user encoding tower by our proposed a moving-aggregation to update user encodings. We have applied our one backpropagation together with a simple projection matrix-based user-item encoder, random negative sampling and plain InfoNCE loss. Despite its rather simple implementation, our OneBP outperforms the state-of-the-art competitors in terms of improved recommendation performance and computation efficiency.

\bibliography{bibfile}

\begin{thebibliography}{10}
\providecommand{\url}[1]{#1}
\csname url@samestyle\endcsname
\providecommand{\newblock}{\relax}
\providecommand{\bibinfo}[2]{#2}
\providecommand{\BIBentrySTDinterwordspacing}{\spaceskip=0pt\relax}
\providecommand{\BIBentryALTinterwordstretchfactor}{4}
\providecommand{\BIBentryALTinterwordspacing}{\spaceskip=\fontdimen2\font plus
\BIBentryALTinterwordstretchfactor\fontdimen3\font minus
  \fontdimen4\font\relax}
\providecommand{\BIBforeignlanguage}[2]{{%
\expandafter\ifx\csname l@#1\endcsname\relax
\typeout{** WARNING: IEEEtran.bst: No hyphenation pattern has been}%
\typeout{** loaded for the language `#1'. Using the pattern for}%
\typeout{** the default language instead.}%
\else
\language=\csname l@#1\endcsname
\fi
#2}}
\providecommand{\BIBdecl}{\relax}
\BIBdecl

\bibitem{Aljukhadar:2012:IJEC}
M.~Aljukhadar, S.~Senecal, and C.-E. Daoust, ``Using recommendation agents to
  cope with information overload,'' \emph{International Journal of Electronic
  Commerce}, vol.~17, no.~2, pp. 41--70, 2012.

\bibitem{Aljukhadar:2010:RecSys}
------, ``Information overload and usage of recommendations,'' in
  \emph{Proceedings of the ACM RecSys 2010 Workshop on User-Centric Evaluation
  of Recommender Systems and Their Interfaces (UCERSTI), Barcelona,
  Spain}.\hskip 1em plus 0.5em minus 0.4em\relax New York, NY, USA: Association
  for Computing Machinery, 2010, pp. 26--33.

\bibitem{Pan:2008:IEEE}
R.~Pan, Y.~Zhou, B.~Cao, N.~N. Liu, R.~Lukose, M.~Scholz, and Q.~Yang,
  ``One-class collaborative filtering,'' in \emph{2008 Eighth IEEE
  international conference on data mining}.\hskip 1em plus 0.5em minus
  0.4em\relax Piscataway: IEEE, 2008, pp. 502--511.

\bibitem{Hu:2008:IEEE}
Y.~Hu, Y.~Koren, and C.~Volinsky, ``Collaborative filtering for implicit
  feedback datasets,'' in \emph{2008 Eighth IEEE international conference on
  data mining}.\hskip 1em plus 0.5em minus 0.4em\relax Piscataway: IEEE, 2008,
  pp. 263--272.

\bibitem{Hsieh:2017:WWW}
C.-K. Hsieh, L.~Yang, Y.~Cui, T.-Y. Lin, S.~Belongie, and D.~Estrin,
  ``Collaborative metric learning,'' in \emph{WWW}.\hskip 1em plus 0.5em minus
  0.4em\relax Republic and Canton of Geneva, CHE: International World Wide Web
  Conferences Steering Committee, 2017, p. 193–201.

\bibitem{Tran:2019:SIGIR}
V.-A. Tran, R.~Hennequin, J.~Royo-Letelier, and M.~Moussallam, ``Improving
  collaborative metric learning with efficient negative sampling,'' in
  \emph{SIGIR}.\hskip 1em plus 0.5em minus 0.4em\relax New York, NY, USA:
  Association for Computing Machinery, 2019, p. 1201–1204.

\bibitem{Liu:2023:ICDE}
B.~Liu and B.~Wang, ``Bayesian negative sampling for recommendation,'' in
  \emph{2023 IEEE 39th International Conference on Data Engineering
  (ICDE)}.\hskip 1em plus 0.5em minus 0.4em\relax Piscataway: IEEE, 2023, pp.
  749--761.

\bibitem{Xiangnan:2020:SIGIR}
H.~Xiangnan, D.~Kuan, W.~Xiang, L.~Yan, Z.~Yongdong, and W.~Meng, ``Lightgcn:
  Simplifying and powering graph convolution network for recommendation.'' in
  \emph{SIGIR}.\hskip 1em plus 0.5em minus 0.4em\relax New York, NY, USA:
  Association for Computing Machinery, 2020, p.~10.

\bibitem{Liang:2018:WWW}
D.~Liang, R.~G. Krishnan, M.~D. Hoffman, and T.~Jebara, ``Variational
  autoencoders for collaborative filtering,'' in \emph{Proceedings of the 2018
  World Wide Web Conference}.\hskip 1em plus 0.5em minus 0.4em\relax New York,
  NY, USA: International World Wide Web Conferences Steering Committee, 2018,
  p. 689–698.

\bibitem{Chae:2018:CIKM}
D.-K. Chae, J.-S. Kang, S.-W. Kim, and J.-T. Lee, ``Cfgan: A generic
  collaborative filtering framework based on generative adversarial networks,''
  in \emph{Proceedings of the 27th ACM International Conference on Information
  and Knowledge Management}.\hskip 1em plus 0.5em minus 0.4em\relax New York,
  NY, USA: Association for Computing Machinery, 2018, p. 137–146.

\bibitem{Li:2022:WWW}
J.~Li, Y.~Ren, and K.~Deng, ``Fairgan: Gans-based fairness-aware learning for
  recommendations with implicit feedback,'' in \emph{Proceedings of the ACM Web
  Conference 2022}.\hskip 1em plus 0.5em minus 0.4em\relax New York, NY, USA:
  Association for Computing Machinery, 2022, p. 297–307.

\bibitem{Bao:2022:NIPS}
S.~Bao, Q.~Xu, Z.~Yang, Y.~He, X.~Cao, and Q.~Huang, ``The minority matters: A
  diversity-promoting collaborative metric learning algorithm,'' \emph{Advances
  in Neural Information Processing Systems}, vol.~35, pp. 2451--2464, 2022.

\bibitem{Bao:2022:IEEE}
S.~Bao, Q.~Xu, Z.~Yang, X.~Cao, and Q.~Huang, ``Rethinking collaborative metric
  learning: Toward an efficient alternative without negative sampling,''
  \emph{IEEE Transactions on Pattern Analysis and Machine Intelligence},
  vol.~45, no.~1, pp. 1017--1035, 2022.

\bibitem{Lee:2021:SIGIR}
D.~Lee, S.~Kang, H.~Ju, C.~Park, and H.~Yu, ``Bootstrapping user and item
  representations for one-class collaborative filtering,'' in
  \emph{SIGIR}.\hskip 1em plus 0.5em minus 0.4em\relax New York, NY, USA:
  Association for Computing Machinery, 2021, p. 317–326.

\bibitem{Li:2020:AAAI}
M.~Li, S.~Zhang, F.~Zhu, W.~Qian, L.~Zang, J.~Han, and S.~Hu, ``Symmetric
  metric learning with adaptive margin for recommendation,'' \emph{Proceedings
  of the AAAI Conference on Artificial Intelligence}, vol.~34, no.~04, pp.
  4634--4641, 2020.

\bibitem{Wei:2023:SIGIR}
T.~Wei, J.~Ma, and T.~W. Chow, ``Collaborative residual metric learning,'' in
  \emph{Proceedings of the 46th International ACM SIGIR Conference on Research
  and Development in Information Retrieval}.\hskip 1em plus 0.5em minus
  0.4em\relax New York, NY, USA: Association for Computing Machinery, 2023, p.
  1107–1116.

\bibitem{Mao:2021:CIKM}
K.~Mao, J.~Zhu, J.~Wang, Q.~Dai, Z.~Dong, X.~Xiao, and X.~He, ``Simplex: A
  simple and strong baseline for collaborative filtering,'' in
  \emph{Proceedings of the 30th ACM International Conference on Information \&
  Knowledge Management}.\hskip 1em plus 0.5em minus 0.4em\relax New York, NY,
  USA: Association for Computing Machinery, 2021, p. 1243–1252.

\bibitem{Huang:2013:CIKM}
P.-S. Huang, X.~He, J.~Gao, L.~Deng, A.~Acero, and L.~Heck, ``Learning deep
  structured semantic models for web search using clickthrough data,'' in
  \emph{Proceedings of the 22nd ACM International Conference on Information \&
  Knowledge Management}.\hskip 1em plus 0.5em minus 0.4em\relax New York, NY,
  USA: Association for Computing Machinery, 2013, p. 2333–2338.

\bibitem{Yang:2020:WWW}
J.~Yang, X.~Yi, D.~Zhiyuan~Cheng, L.~Hong, Y.~Li, S.~Xiaoming~Wang, T.~Xu, and
  E.~H. Chi, ``Mixed negative sampling for learning two-tower neural networks
  in recommendations,'' in \emph{Companion Proceedings of the Web Conference
  2020}.\hskip 1em plus 0.5em minus 0.4em\relax New York, NY, USA: Association
  for Computing Machinery, 2020, p. 441–447.

\bibitem{Su:2023:SIGIR}
L.~Su, F.~Yan, J.~Zhu, X.~Xiao, H.~Duan, Z.~Zhao, Z.~Dong, and R.~Tang,
  ``Beyond two-tower matching: Learning sparse retrievable cross-interactions
  for recommendation,'' in \emph{Proceedings of the 46th International ACM
  SIGIR Conference on Research and Development in Information Retrieval}.\hskip
  1em plus 0.5em minus 0.4em\relax New York, NY, USA: Association for Computing
  Machinery, 2023, p. 548–557.

\bibitem{Chen:2020:ICML}
T.~Chen, S.~Kornblith, M.~Norouzi, and G.~Hinton, ``A simple framework for
  contrastive learning of visual representations,'' in \emph{International
  conference on machine learning}.\hskip 1em plus 0.5em minus 0.4em\relax New
  York, NY, USA: Association for Computing Machinery, 2020, pp. 1597--1607.

\bibitem{Chuang:2020:NIPS}
C.-Y. Chuang, J.~Robinson, Y.-C. Lin, A.~Torralba, and S.~Jegelka, ``Debiased
  contrastive learning,'' \emph{Advances in Neural Information Processing
  Systems}, vol.~33, pp. 8765--8775, 2020.

\bibitem{Khosla:2020:NIPS}
P.~Khosla, P.~Teterwak, C.~Wang, A.~Sarna, Y.~Tian, P.~Isola, A.~Maschinot,
  C.~Liu, and D.~Krishnan, ``Supervised contrastive learning,'' \emph{Advances
  in neural information processing systems}, vol.~33, pp. 18\,661--18\,673,
  2020.

\bibitem{Song:2020:NIPS}
J.~Song and S.~Ermon, ``Multi-label contrastive predictive coding,''
  \emph{Advances in neural information processing systems}, vol.~33, pp.
  8161--8173, 2020.

\bibitem{Xiangnan:2017:WWW}
X.~He, L.~Liao, H.~Zhang, L.~Nie, X.~Hu, and T.-S. Chua, ``Neural collaborative
  filtering,'' in \emph{WWW}.\hskip 1em plus 0.5em minus 0.4em\relax Republic
  and Canton of Geneva, CHE: International World Wide Web Conferences Steering
  Committee, 2017, p. 173–182.

\bibitem{Wang:2019:SIGIR}
X.~Wang, X.~He, M.~Wang, F.~Feng, and T.-S. Chua, ``Neural graph collaborative
  filtering,'' in \emph{SIGIR}.\hskip 1em plus 0.5em minus 0.4em\relax New
  York, NY, USA: Association for Computing Machinery, 2019, p. 165–174.

\bibitem{Chen:2017:KDD}
T.~Chen, Y.~Sun, Y.~Shi, and L.~Hong, ``On sampling strategies for neural
  network-based collaborative filtering,'' in \emph{Proceedings of the 23rd ACM
  SIGKDD International Conference on Knowledge Discovery and Data
  Mining}.\hskip 1em plus 0.5em minus 0.4em\relax New York, NY, USA:
  Association for Computing Machinery, 2017, p. 767–776.

\bibitem{Steffen:2009:UAI}
S.~Rendle, C.~Freudenthaler, Z.~Gantner, and L.~Schmidt{-}Thieme, ``Bpr:
  Bayesian personalized ranking from implicit feedback,'' in \emph{UAI 2009,
  Proceedings of the Twenty-Fifth Conference on Uncertainty in Artificial
  Intelligence, Montreal, QC, Canada, June 18-21, 2009}.\hskip 1em plus 0.5em
  minus 0.4em\relax Arlington, Virginia, USA: AUAI Press, 2009, pp. 452--461.

\bibitem{Zhang:2013:SIGIR}
W.~Zhang, T.~Chen, J.~Wang, and Y.~Yu, ``Optimizing top-n collaborative
  filtering via dynamic negative item sampling,'' in \emph{Proceedings of the
  36th International ACM SIGIR Conference on Research and Development in
  Information Retrieval}.\hskip 1em plus 0.5em minus 0.4em\relax New York, NY,
  USA: Association for Computing Machinery, 2013, pp. 785--788.

\bibitem{Chen:2023:WWW}
X.~Chen, W.~Fan, J.~Chen, H.~Liu, Z.~Liu, Z.~Zhang, and Q.~Li, ``Fairly
  adaptive negative sampling for recommendations,'' in \emph{Proceedings of the
  ACM Web Conference 2023}.\hskip 1em plus 0.5em minus 0.4em\relax New York,
  NY, USA: Association for Computing Machinery, 2023, p. 3723–3733.

\bibitem{Oord:2018:arxiv}
A.~v.~d. Oord, Y.~Li, and O.~Vinyals, ``Representation learning with
  contrastive predictive coding,'' \emph{arXiv preprint arXiv:1807.03748},
  2018.

\bibitem{Koren:2009:Computer}
Y.~Koren, R.~Bell, and C.~Volinsky, ``Matrix factorization techniques for
  recommender systems,'' \emph{Computer}, vol.~42, no.~8, pp. 30--37, 2009.

\bibitem{Mikolov:2013:NIPS}
T.~Mikolov, I.~Sutskever, K.~Chen, G.~Corrado, and J.~Dean, ``Distributed
  representations of words and phrases and their compositionality,''
  \emph{Advances in neural information processing systems}, vol.~26, pp.
  3111--3119, 2013.

\bibitem{Tang:2015:WWW}
J.~Tang, M.~Qu, M.~Wang, M.~Zhang, J.~Yan, and Q.~Mei, ``Line: Large-scale
  information network embedding,'' in \emph{Proceedings of the 24th
  International Conference on World Wide Web}.\hskip 1em plus 0.5em minus
  0.4em\relax Republic and Canton of Geneva, CHE: International World Wide Web
  Conferences Steering Committee, 2015, pp. 1067--1077.

\bibitem{Liu:2023:arxiv}
B.~Liu, E.~Chen, and B.~Wang, ``Reducing popularity bias in recommender systems
  through auc-optimal negative sampling,'' \emph{arXiv preprint
  arXiv:2306.01348}, 2023.

\bibitem{Yu:2018:CIKM}
R.~Yu, Y.~Zhang, Y.~Ye, L.~Wu, C.~Wang, Q.~Liu, and E.~Chen, ``Multiple
  pairwise ranking with implicit feedback,'' in \emph{Proceedings of the 27th
  ACM Int. Conference on Information and Knowledge Management, ACM}.\hskip 1em
  plus 0.5em minus 0.4em\relax New York, NY, USA: Association for Computing
  Machinery, 2018, pp. 1727--1730.

\bibitem{Robinson:2020:ICLR}
J.~D. Robinson, C.-Y. Chuang, S.~Sra, and S.~Jegelka, ``Contrastive learning
  with hard negative samples,'' in \emph{International Conference on Learning
  Representations}.\hskip 1em plus 0.5em minus 0.4em\relax New York, NY, USA:
  OpenReview.net, 2020, p. 2801.

\bibitem{Rottmann:2020:EDA}
M.~Rottmann, K.~Maag, R.~Chan, F.~H\"{u}ger, P.~Schlicht, and H.~Gottschalk,
  ``Detection of false positive and false negative samples in semantic
  segmentation,'' in \emph{Proceedings of the 23rd Conference on Design,
  Automation and Test in Europe}.\hskip 1em plus 0.5em minus 0.4em\relax San
  Jose, CA, USA: EDA Consortium, 2020, p. 1351–1356.

\bibitem{Li:2023:ACM}
H.~Li, Y.~Bin, J.~Liao, Y.~Yang, and H.~T. Shen, ``Your negative may not be
  true negative: Boosting image-text matching with false negative
  elimination,'' in \emph{Proceedings of the 31st ACM International Conference
  on Multimedia}.\hskip 1em plus 0.5em minus 0.4em\relax New York, NY, USA:
  Association for Computing Machinery, 2023, p. 924–934.

\bibitem{Chen:2021:ICLR}
T.-S. Chen, W.-C. Hung, H.-Y. Tseng, S.-Y. Chien, and M.-H. Yang, ``Incremental
  false negative detection for contrastive learning,'' in \emph{International
  Conference on Learning Representations}.\hskip 1em plus 0.5em minus
  0.4em\relax New York, NY, USA: OpenReview.net, 2021.

\bibitem{Ding:2020:NIPS}
J.~Ding, Y.~Quan, Q.~Yao, Y.~Li, and D.~Jin, ``Simplify and robustify negative
  sampling for implicit collaborative filtering,'' \emph{Advances in Neural
  Information Processing Systems}, vol.~33, pp. 1094--1105, 2020.

\bibitem{Wang:2020:ACM}
X.~Wang, Y.~Xu, X.~He, Y.~Cao, M.~Wang, and T.-S. Chua, ``Reinforced negative
  sampling over knowledge graph for recommendation,'' in \emph{Proceedings of
  The Web Conference 2020}.\hskip 1em plus 0.5em minus 0.4em\relax New York,
  NY, USA: Association for Computing Machinery, 2020, p. 99–109.

\bibitem{Qin:2020:AAAI}
X.~Qin, N.~Sheikh, B.~Reinwald, and L.~Wu, ``Relation-aware graph attention
  model with adaptive self-adversarial training,'' \emph{Proceedings of the
  AAAI Conference on Artificial Intelligence}, vol.~35, no.~11, pp. 9368--9376,
  2021.

\bibitem{Neyman:1967:Berkeley}
J.~Neyman and E.~L. SCOTT, ``Berkeley symposium on mathematical statistics and
  probability,'' in \emph{Proceedings of the Berkeley Symposium on Mathematical
  Statistics and Probability, eds LM Le Cam and J. Neyman (Berkeley, CA:
  University of California Press)}, 1967.

\bibitem{Laurens:2008:JMLR}
L.~van~der Maaten and G.~E. Hinton, ``Visualizing data using t-sne,''
  \emph{Journal of Machine Learning Research}, vol.~9, pp. 2579--2605, 2008.

\end{thebibliography}

\end{document}